\documentclass[twocolumn,pra,floatfix,showpacs,superscriptaddress,amssymb,amsmath,amsmath]{revtex4-2}
\usepackage{graphicx}
\usepackage{epstopdf}
\usepackage{comment}
\usepackage{hyperref}
\usepackage{amsmath}
\usepackage{bm}
\usepackage{adjustbox}
\usepackage{tablefootnote}
\usepackage{times}
\usepackage[dvipsnames]{xcolor}

\usepackage{dcolumn}
\newcolumntype{d}[1]{D{.}{.}{#1}}

\hypersetup{pdfstartpage=1,pdfmenubar=true,pdfpagemode=UseNone,pdftoolbar=true,
urlcolor=blue,linkcolor=blue,colorlinks = true,citecolor=blue, bookmarksopen=false}

\newcommand\mydots{\hbox to 1em{.\hss.\hss.}}
\newcommand{\stX}{$X^2\Sigma^+_u$}
\newcommand{\stA}{$A^2\Sigma^+_g$}
\newcommand{\sta}{$a^4\Sigma^+_u$} 
\newcommand{\stB}{$B^2\Sigma^+_g$}
\newcommand{\stC}{$C^2\Sigma^+_u$}
\newcommand{\stb}{$b^4\Sigma^+_g$}

\begin{document}
\title{\emph{Ab initio} potential energy curves, scattering lengths, and rovibrational levels of the He$_2^+$ molecular ion in excited electronic states} 
\author{Jacek G\ifmmode\mbox{\c{e}}\else\c{e}\fi{}bala}
\affiliation{Faculty of Physics, University of Warsaw, Pasteura 5, 02-093 Warsaw, Poland}
\author{Micha\l{} Przybytek}
\email{m.przybytek@uw.edu.pl}
\affiliation{Faculty of Chemistry, University of Warsaw, Pasteura 1, 02-093 Warsaw, Poland}
\author{Marcin Gronowski}
\affiliation{Faculty of Physics, University of Warsaw, Pasteura 5, 02-093 Warsaw, Poland}
\author{Micha{\l} Tomza}
\email{michal.tomza@fuw.edu.pl}
\affiliation{Faculty of Physics, University of Warsaw, Pasteura 5, 02-093 Warsaw, Poland}

\date{\today}

\begin{abstract}
We calculate accurate potential energy curves for a ground-state He$^+$ ion interacting with a He atom in the lowest-energy metastable $^3\!S$ electronic state. We employ the full configuration interaction method, equivalent to exact diagonalization, with results extrapolated to the complete basis set limit. The leading relativistic and adiabatic corrections are included using perturbation theory. We calculate rovibrational levels and spectroscopic constants of the He$_2^+$ molecular ion in excited electronic states for three stable isotopologues. We predict the scattering lengths for ultracold ion-atom collisions. The theoretical data are presented with their uncertainties and agree well with previous results for the ground state. The reported results may be useful for the spectroscopy of the He$_2^+$ molecular ion in the excited electronic state and collisional studies of He$^+$ ions immersed in ultracold gases of metastable He atoms.
\end{abstract}

\maketitle

\section{Introduction}

Precision spectroscopy of few-electron atoms and molecules provides a perfect ground for probing our understanding of quantum mechanics, quantum electrodynamics, and even the Standard Model~\cite{HanschRMP06,SafronovaRMP18}. Examples include the measurements of the atomic hydrogen Lamb shift and the proton radius~\cite{PohlNature10,BezginovScience19}, the electron-to-proton mass ratio~\cite{AlighanbariNature20}, the spatio-temporal variation of fundamental constants~\cite{SchillerPRL14}, and bounds on the fifth force~\cite{SalumbidesPRD13}, among others. Such applications are realized by comparing experimental atomic or molecular transition frequencies with theoretical results of accurate \emph{ab initio} calculations.

The electronic structure calculations for few-electron atoms and molecules have reached unparalleled accuracy. The non-relativistic, electronic Schr\"odinger equation can be solved almost exactly for such systems, while relativistic, quantum electrodynamics, adiabatic, and nonadiabatic corrections can be included in a controlled and systematic way. Highly accurate calculations have been carried out for the H~\cite{yerokhin_pachucki_He_2018}, He~\cite{LachPRA04, Pachucki_He_dokladny,PuchalskiPRA20}, and Li~\cite{Pachucki_Lit_2010, Pachucki_Lit_2011}  atoms, as well as for the H$_2^+$~\cite{PhysRevA.77.022509}, H$_2$~\cite{PuchalskiPRL19}, HeH$^+$~\cite{KolosCP76,BishopJMS79}, HeH~\cite{Lo_HeH_2006,Burrows_2017}, He$^+_2$~\cite{TungJCP12,JansenPRL18,SemeriaPRL20,FerencPRL20}, and He$_2$~\cite{PrzybytekPRL10,PrzybytekPRL17} molecules and molecular ions, and their isotopologues, reaching accuracy below $10^{-4}\,$cm$^{-1}$ or 1 ppb in some cases. Spectroscopic accuracy (below $1\,$cm$^{-1}$) has also been achieved for some larger, many-electron molecules, e.g., Li$_2$~\cite{LesiukPRA20}, NaLi~\cite{GronowskiPRA20}, and Be$_2$~\cite{LesiukJCTC19}. The precisely calculated spectroscopic properties of the lightest atoms and molecules are also important for astrochemistry and astrophysics, as they dominate the cosmic matter~\cite{herbst_2001,FortenberryMA20}.

The highly accurate calculations of interactions between atoms are crucial for predicting and understanding their collisional properties in the ultracold regime. Tremendous successes have already been achieved in experimental control and application of degenerate quantum gases of ultracold neutral atoms~\cite{RevModPhys.82.1225}. Recently, the first experiments with single trapped ions immersed in ultracold atomic gases~\cite{feldker_2020,weckesser2021} have also achieved the $s$-wave quantum regime of ion-atom collisions. It paves the way for investigating qualitatively new and interesting controlled chemistry and quantum few-body physics~\cite{TomzaRMP19} governed by longer-range ion-atom interactions, which are missing in neutral samples. Until recently, the experimental ion-atom systems were only investigated in the temperature regime where the semiclassical theory of collisions explains the system's dynamics~\cite{PhysRevA.57.906}. It is easier to reach the quantum regime in systems that contain a heavy ion and light atoms, where the micromotion effects are less relevant and thus result in less heating~\cite{CetinaPRL12}. On the other hand, one can investigate ion-atom interactions in light systems, where the small reduced mass of the colliding particles results in a higher temperature of the quantum regime~\cite{SchmidPRL18}. Additionally, a small number of electrons may allow for more accurate theoretical predictions. We choose the latter approach and propose studying a helium ion in the ground doublet $^2\!S$ state immersed in an ultracold gas of helium atoms in the first excited metastable triplet $^3\!S$ state. The well-developed methods for laser cooling and trapping of such atoms~\cite{TrippenbachMetastableRev2012} combined with high controllability and long lifetime of ions, which can be sympathetically cooled and manipulated~\cite{KatzNP22}, give the prospects for fruitful experimental investigation and application of ultracold He$^+$+He$^*$ collisions.  

Here, we investigate the excited-electronic-state properties of the He$_2^+$ molecular ion resulting from the interaction between a He$^+$($^2\!S$) ion and a He$^*$($^3\!S$) atom. We use state-of-the-art electronic structure methods to calculate the potential-energy curves (PECs) and spectroscopic constants. We employ the full configuration interaction (FCI) method and one-electron Gaussian basis sets of increasing size. The electronic energies near the complete basis set (CBS) limit are obtained by employing extrapolation techniques. In order to increase the accuracy of the calculations, we include the leading relativistic and adiabatic corrections using perturbation theory and also estimate the quantum electrodynamics effects on the interaction energy. We calculate the rovibrational levels of three stable isotopologues of the He$_2^+$ molecular ion. Finally, we use our accurate PECs to provide the scattering lengths in the ground and excited electronic states.

This article has the following structure. Section~\ref{sec:theory} describes the theoretical methods used in the \emph{ab initio} electronic structure and nuclear dynamics calculations. Section~\ref{sec:results} presents and discusses the results, i.e., the potential energy curves, the rovibrational spectra, the spectroscopic characteristics of the He$_2^+$ molecular ion, and the scattering lengths. Section~\ref{sec:summary} summarizes the paper and points to further applications and extensions of the presented results and methodology.

\section{Computational methods}
\label{sec:theory}

\subsection{Electronic structure calculations}
\label{sec:ES}

The interaction of an open-shell helium ion in the ground doublet $^2\!S$ state with a closed-shell helium atom in the ground singlet $^1\!S$ electronic state results in the ground molecular electronic state of the \stX{} symmetry and the first-excited electronic state of the \stA{} symmetry. When a helium ion in the ground $^2\!S$ state interacts with an open-shell helium atom in the first metastable triplet $^3\!S$ state, it results in the electronic states of the \sta, \stB, \stC, and \stb{} symmetries, listed in the order of increasing energy. We solve the electronic time-independent non-relativistic Schrödinger equation in the Born-Oppenheimer approximation and obtain wavefunctions and energies of the mentioned molecular electronic states. 

The many-electron wavefunctions are represented using the configuration interaction (CI) approach~\cite{CI_method_1999} as an expansion in a set of Slater determinants constructed from Hartree-Fock spinorbitals, which are in turn expanded in a set of fixed one-electron Gaussian basis functions. The energies and wavefunctions of the considered electronic states are obtained by the direct diagonalization of the Hamiltonian matrix calculated using all possible Slater determinants with a well-defined spatial symmetry and spin projection: $M_S=\frac12$ for the doublet molecular states, $M_S=\frac32$ for the quartet molecular states, $M_S=0$ for the singlet atomic states, $M_S=\frac12$ for the doublet ionic states, and $M_S=1$ for the triplet atomic states. The combination of the CI approach and the direct diagonalization in a set of all possible Slater determinants is known as the full CI (FCI) method. With modern computing power, highly accurate calculations employing the FCI method are possible for systems comprising up to 8 electrons~\cite{OLSEN1990463, HIRAI2022111460, Kohn2022}.

We calculate the non-relativistic Born-Oppenheimer (BO) interaction energy at the internuclear distance $R$ using the super-molecule method
\begin{equation}
V_\text{BO}(R) = E_\text{BO}^{\text{He}\text{He}^+}\!(R) 
- E_\text{BO}^{\text{He}}(R) - E_\text{BO}^{\text{He}^+}\!(R) \ ,
\label{definicjaFCI}
\end{equation}
where $E_\text{BO}^{\text{He}\text{He}^+}\!(R)$ is the BO energy of a given molecular state and $E_\text{BO}^{\text{He}}(R)$ and $E_\text{BO}^{\text{He}^+}\!(R)$ are the BO energies of the atom (in the respective electronic state) and the ion computed in the dimer-centered basis set~\cite{DCBS95}. This is equivalent to applying the so-called counterpoise scheme~\cite{Boys1970} to remove the basis set superposition error (BSSE), which is a consequence of unphysical lowering of the monomer energies due to the presence of basis functions at both sites in calculations for the dimer. When $R \rightarrow \infty$, the energies $E_\text{BO}^{\text{He}}(R)$ and $E_\text{BO}^{\text{He}^+}\!(R)$ tend to values $E_\text{BO}^{\text{He}}$ and $E_\text{BO}^{\text{He}^+}$, which are the energies of the noninteracting atom and ion, respectively, in the monomer-centered basis set. 

To achieve the highest accuracy, we include the adiabatic and relativistic corrections to the interaction energy
\begin{equation}
V_\text{int}(R) =  V_\text{BO}(R) + V_\text{ad}(R) + V_\text{rel}(R) \ .
\label{eq:Vint}
\end{equation}

The adiabatic correction, also known as the diagonal Born-Oppenheimer correction, is the leading effect beyond the Born-Oppenheimer approximation due to the coupling of nuclear and electronic motions. In light systems, such as  He$_2^+$, the adiabatic correction may be dominant when compared with the relativistic correction, especially at smaller internuclear distances. The adiabatic correction to the interaction energy, $V_\text{ad}(R)$, is constructed analogously to Eq.~\eqref{definicjaFCI} with the BO energies, $E_\text{BO}(R)$, of the molecular, atomic, and ionic states replaced by the adiabatic correction, $E_\text{ad}(R)$, calculated for each respective system. Within the Born-Handy approach~\cite{Handy:86,Ioannou:96,Handy:96}, the adiabatic correction for a given system is defined formally as the expectation value of the nuclear kinetic energy operator, $\langle T_n \rangle$, calculated with the non-relativistic Born-Oppenheimer wavefunction in the space-fixed coordinate space~\cite{Kutzelnigg:97}. As the BO wavefunction depends on the nuclear coordinates only parametrically, direct implementation of this definition requires the use of cumbersome numerical differentiation techniques~\cite{KOMASA1999293,Valeev2003}. Fully analytical approaches to calculate the adiabatic correction have also been developed~\cite{Jensen1988,Tajti2006,Tajti2007}. In this work, we employ an alternative approach designed specifically for diatomic systems proposed in Ref.~\cite{NAPT08} and described in detail in the Appendix. The presented approach relies on the FCI representation of the wavefunction of the dimer and the monomers.

We calculate the leading-order relativistic correction to the interaction energy, $V_\text{rel}(R)$, within perturbation theory formalism. This approximation is appropriate because we only investigate the $\Sigma$ states of light molecules involving atoms with small atomic numbers. We employ the approach based on the Breit-Pauli Hamiltonian~\cite{Bethe1957} accurate to the second order in the fine structure constant ($\sim \alpha^2$)
\begin{align}
\label{breit}
V_\text{rel}(R)&=V_\text{MV}(R) +V_\text{D1}(R) + V_\text{D2}(R) + V_\text{OO}(R) \ ,
\end{align}
where $V_\text{MV}(R)$ is the mass-velocity correction, $V_\text{D1}(R)$ and $V_\text{D2}(R)$ are the one- and two-electron Darwin corrections, and $V_\text{OO}(R)$ is the orbit-orbit correction. We neglect the spin-spin and spin-orbit corrections that result in the fine-structure splitting of the energy levels and include only terms that shift non-relativistic energy without splitting the levels. The individual relativistic corrections in Eq.~\eqref{breit} are obtained analogously to Eq.~\eqref{definicjaFCI} from corrections calculated for the molecular, atomic, and ionic states. The relativistic corrections to the energies of the molecule and the monomers, in atomic units, are given by expectation values calculated using the FCI wavefunctions
\begin{align}
    \label{p4}
    E_\text{MV}(R) &= -\frac{\alpha^2}{8} \left\langle \sum_i \textbf{p}_i^4 \right\rangle,\\
    \label{d1}
    E_\text{D1}(R) &= \frac{\pi}{2}\alpha^2\sum_I Z_I \left\langle \sum_i 
    \delta(\textbf{r}_{Ii})\right\rangle,\\
    \label{d2}
    E_\text{D2}(R) &= -\pi\alpha^2\left\langle\sum_{i>j}\delta(\textbf{r}_{ij})\right\rangle,\\
    \label{bb}
    E_\text{OO}(R)&=  -\frac{\alpha^2}{2}\left\langle \sum_{i>j} 
    \left[\frac{\textbf{p}_i\cdot\textbf{p}_j}{r_{ij}}
    +\frac{\textbf{r}_{ij}\cdot(\textbf{r}_{ij}\cdot\textbf{p}_j)\textbf{p}_i}{r_{ij}^3}\right] \right\rangle
    ,
\end{align}
where $I$ and $i$ denote the nuclei and electrons, respectively, $Z_I$ is the atomic number of a given nucleus, $\textbf{r}_{Ii}$ and $\textbf{r}_{ij}$ denote interparticle vectors, $\delta\left(\textbf{r}\right)$ is the Dirac-delta function, 
and $\textbf{p}_i$ is the momentum operator of the $i$-th electron.

Additionally, we estimate the leading-order post-Breit-Pauli correction to the interaction energy, $V_\text{QED}(R)$. These terms, called the Lamb shifts, are proportional to $\alpha^3$ and $\alpha^3\ln\alpha$, and estimate the leading quantum electrodynamics (QED) effects. They are given by~\cite{Caswell_Lepage_1986}
\begin{equation}
V_\text{QED}(R) = V_{\text{QED},1}(R) + V_{\text{QED},2}(R) \ ,
\end{equation}
where $V_{\text{QED},1}(R)$ and $V_{\text{QED},2}(R)$ are the one- and two-electron contributions approximated by
\begin{align}
V_{\text{QED},1}(R) & = \frac{8\alpha}{3\pi}
\left(\frac{19}{30}-2\ln \alpha - \ln k_0^{\text{HeHe}^+}\right) V_\text{D1}(R) \ ,
\\
V_{\text{QED},2}(R) &= -\frac{\alpha}{\pi} 
\left(\frac{89}{15}+\frac{14}{3}\ln \alpha\right) V_\text{D2}(R) \ ,
\end{align}
where $V_\text{D1}(R)$ and $V_\text{D2}(R)$ are the one- and two-electron Darwin corrections to the interaction energy from Eq.~\eqref{breit} obtained using Eqs.~\eqref{d1} and \eqref{d2}. Furthermore, we approximate the molecular Bethe logarithm by
\begin{equation}
\ln k_0^{\text{HeHe}^+}=\frac
{\ln k_0^\text{He}E_\text{D1}^\text{He}+\ln k_0^{\text{He}^+}E_\text{D1}^{\text{He}^+}}
{E_\text{D1}^\text{He}+E_\text{D1}^{\text{He}^+}} \ ,
\label{meanBethe}
\end{equation}
neglecting its dependence on $R$, where $E^{\text{He}}_\text{D1}$ and $E^{\text{He}^+}_\text{D1}$ are the one-electron Darwin corrections to the energy of monomers and $\ln k_0^{\text{He}}$ and $\ln k_0^{\text{He}^+}$ are the atom and ion Bethe logarithms, respectively~\cite{Piszczatowski_2009}. We use $\ln k_0^\text{He}=4.3701602230703(3)$~\cite{Korobov_2019}, $\ln k_0^{\text{He}^*}= 4.364036820476(1)$~\cite{Korobov_2019}, and $\ln k_0^{\text{He}^+}=4.370422916$ (compiled from Ref.~\cite{Drake_Goldman_2000}). 

The FCI calculations are performed using the \textsc{Hector} program~\cite{hector}. The Hartree-Fock orbitals, the standard one- and two-electron integrals, nonstandard integrals necessary in the calculation of the adiabatic correction, and integrals involving the relativistic operators are generated using local version of the \textsc{Dalton} 2.0 package~\cite{Dalton2020,dalton2,Coriani:04}. We calculate the values of $V_\text{BO}(R)$ on a grid of 101 points of $R$ ranging from $1.2$ to $50.0$~a$_0$ for the \stX{} state and 79 points of $R$ in the same range for all other molecular electronic states~\cite{supplemental}. We use a grid step of $0.25$~a$_0$ near the minima of the potentials ($0.05$~a$_0$ for \stX{}) and $0.50-1.00$~a$_0$ elsewhere. All relativistic and the adiabatic corrections are calculated on a grid of 43 points of $R$ ranging from $1.25$ to $50.0$~a$_0$~\cite{supplemental}. The interaction energies are interpolated using spline polynomials of the order 3 on a dense grid instead of fitting a single analytical function in order to avoid potential fitting errors.

In the FCI calculations of the molecular electronic states dissociating into the helium atom in the singlet $^1\!S$ state and the helium ion in the doublet $^2\!S$ state, we use a family of double-augmented correlation-consistent polarized-valence Gaussian basis sets d$X$Z($^1\!S$) from Ref.~\cite{cencek_2012}, where the cardinal number $X$ ranges from $X=2$ to $X=8$. The largest angular momentum of the one-electron functions included in this type of basis sets is $l_\text{max}=X-1$. For the relativistic and adiabatic corrections to the \stX{} and \stA{} states, we use the d$X$Zu($^1\!S$)~\cite{cencek_2012} and d$X$Zcp($^1\!S$)~\cite{PrzybytekPRL17}, respectively, modified versions of the d$X$Z($^1\!S$) basis sets.

In the FCI calculations of the molecular electronic states dissociating into the helium atom in the triplet $^3\!S$ state and the helium ion in the doublet $^2\!S$ state, we use a family of double-augmented correlation-consistent polarized-valence Gaussian basis sets d$X$Z($^3\!S$) from Ref.~\cite{przybytek_jeziorski_2005}, where the cardinal number $X$ ranges from $X=2$ to $X=7$. For the relativistic and adiabatic corrections to the \sta, \stB, \stC, and \stb{} states, we use the d$X$Zu($^3\!S$) and d$X$Zcp($^3\!S$) variants of these basis sets, which are prepared similarly as in the case of the states dissociating into the lowest asymptote. 

The transition electric-dipole moments for electric-dipole-allowed transitions $E1$ are calculated as transition matrix elements between the FCI wavefunctions calculated using the d$X$Z($^3\!S$) basis sets. Additionally, the electric-dipole-forbidden transition moments between doublet and quartet molecular states are estimated by solving the four-component problem with the Dirac-Coulomb Hamiltonian by using the Generalized Active Space CI module~\cite{OLSEN1990463,doi:10.1063/1.1590636,doi:10.1063/1.2176609,doi:10.1063/1.3276157} of the \textsc{Dirac} 2022 program~\cite{DIRAC22}, where we use the doubly augmented Dyall quadruple-zeta basis set~\cite{HeDyall} and include 48 orbitals in the active space.

\subsection{Numerical uncertainties}
\label{sec:uncertainties}

The accuracy of the interaction energies is crucial for calculating rovibrational spectra and scattering lengths. To estimate the accuracy of the calculated interaction energies, we approximate exact values of the interaction energy by extrapolating the results obtained with finite-size basis sets to the complete basis set (CBS) limit and estimate the numerical uncertainties by comparing the results computed with the largest basis set used in the extrapolation procedure with the CBS limit. 

For the doublet molecular electronic states, we assume that the basis set truncation error of the calculated BO interaction energy vanishes with the increasing value of the basis set's cardinal number as~\cite{Halkier1998,Helgaker2008,Klopper2001}
\begin{equation}
V_\text{BO}^X = V_\text{BO}^\text{CBS} + \frac{A_3}{X^3} + \frac{A_5}{X^5} \ ,
\label{extrapolation:dublet}
\end{equation}
where $V_\text{BO}^X$ is the BO interaction energy calculated using basis set with the cardinal number $X$ and $V_\text{BO}^\text{CBS}$ is the BO interaction energy in the CBS limit. We calculate $V_\text{BO}^\text{CBS}$ directly by solving a set of linear equations obtained by applying Eq.~\eqref{extrapolation:dublet} to the results calculated with the three largest available values of $X$: $X=6,7,8$ for the \stX{} and \stA{} states where the  d$X$Z$(^1\!S)$ basis sets are used, and $X=5,6,7$ for the \stB{} and \stC{} states where the d$X$Z$(^3\!S)$ basis sets are used. We then estimate the numerical uncertainty of the BO interaction energy for these states by $\delta V_\text{BO} = |V_\text{BO}^8 - V^\text{CBS}_\text{BO}|$ and $\delta V_\text{BO} = |V_\text{BO}^7 - V^\text{CBS}_\text{BO}|$, respectively.

In the case of fully spin-polarized two-electron triplet and three-electron quartet systems considered in this work, the electronic wavefunction of the system does not involve singlet electron pairs. Therefore, the energy of the metastable helium atom, the energies of the molecular quartet states, and, consequently, the BO interaction energies for these states are expected to converge faster to the CBS limit with the basis set's cardinal number than for the molecular doublet states~\cite{Kutzelnigg1992}. We assume that the basis set truncation error of the calculated BO interaction energy for the \sta{} and \stb{} states vanishes with the increasing value of $X$ as~\cite{Klopper2001}
\begin{equation}
V_\text{BO}^X = V_\text{BO}^\text{CBS} + \frac{A_5}{X^5} \ ,
\label{extrapolation:quartet}
\end{equation}
where $V_\text{BO}^X$ is the BO interaction energy calculated with the d$X$Z$(^3\!S)$ basis set and $V_\text{BO}^\text{CBS}$ is the BO interaction energy in the CBS limit. We calculate $V_\text{BO}^\text{CBS}$ from a two-point formula that can be derived by applying Eq.~\eqref{extrapolation:quartet} to the results obtained with the two largest values of $X$ in the d$X$Z$(^3\!S)$ family, $X=6,7$. The numerical uncertainty of the BO interaction energy for the molecular quartet states is then estimated by $\delta V_\text{BO} = |V_\text{BO}^7 - V^\text{CBS}_\text{BO}|$.

We also perform the CBS extrapolations of the adiabatic correction to the interaction energy and all individual components of the relativistic correction in Eq.~\eqref{breit}. Additionally, we extrapolate separately a sum of the mass-velocity and one-electron Darwin terms,
\begin{equation}
V_\text{CG}(R) = V_\text{MV}(R) + V_\text{D1}(R)\,,
\end{equation}
known as the Cowan-Griffin correction~\cite{Cowan:76}. The one-electron relativistic corrections are defined through highly singular operators, see Eqs.~\eqref{p4} and \eqref{d1}, and require a proper description of the electronic wavefunction near the electron-nucleus coalescence points. This requirement is difficult to fulfill in calculations employing Gaussian basis sets, as the Gaussian $s$ orbitals do not exhibit a cusp when the electron-nucleus distance goes to zero. The mass-velocity and one-electron Darwin corrections usually have opposite signs what allows for partial cancellation of errors when they are summed together~\cite{Piszczatowski2008}. Therefore, the results obtained from the extrapolation of the Cowan-Griffin correction are expected to be more accurate than the results calculated as a sum of both components extrapolated separately~\cite{cencek_2012}.

For the post-BO corrections we use a common extrapolation function of the form
\begin{equation}
V_Y^X = V^\text{CBS}_Y + \frac{A}{X^{n_Y}} \ ,
\label{extrapolationCorrection}
\end{equation}
where $Y$ = ad, MV, D1, D2, OO, or CG. $V_Y^X$ represents the correction $Y$ calculated using basis set with the cardinal number $X$, $V_Y^\text{CBS}$ is the value of this correction in the CBS limit, and the exponent $n_Y$ depends on both the correction $Y$ and the multiplicity of the molecular state. The values of $n_Y$ are discussed in Sec.~\ref{sec:monomer}. 
We determine the CBS-limits, $V_Y^\text{CBS}$, from a two-point formula derived by applying Eq.~\eqref{extrapolationCorrection} to the results obtained using basis sets described in Sec.~\ref{sec:ES} with cardinal numbers $X=5$ and $X=6$. We then estimate the uncertainties of the corrections by $\delta V_Y = |V_Y^6 - V_Y^\text{CBS}|$.

Using the procedure described above, we obtain directly the extrapolated values of the adiabatic correction and estimations of its uncertainty. By contrast, the final results for the CBS limit of the relativistic correction is calculated as a sum of independently extrapolated Cowan-Griffin, two-electron Darwin, and orbit-orbit corrections, $V_\text{rel}^\text{CBS}= V_\text{CG}^\text{CBS} + V_\text{D2}^\text{CBS} + V_\text{OO}^\text{CBS}$, and the final estimation of the numerical uncertainty of the relativistic correction is obtained by adding in quadrature the uncertainties of the components,
\begin{equation}
\delta V_\text{rel}=\sqrt{
 \left(\delta V_\text{CG}\right)^2 
+\left(\delta V_\text{D2}\right)^2
+\left(\delta V_\text{OO}\right)^2
} \ .
\end{equation}

We perform the CBS extrapolations for all values of the internuclear distance $R$ separately to obtain final recommended values of the components of the total interaction energy, $V_\text{BO}(R)$, $V_\text{ad}(R)$, and $V_\text{rel}(R)$, together with their estimated numerical uncertainties, $\delta V_\text{BO}(R)$, $\delta V_\text{ad}(R)$, and $\delta V_\text{rel}(R)$, respectively. Next, we calculate the considered rovibrational and scattering quantities using a PEC with the added [$V_\text{int}^{+}(R)$] or subtracted [$V_\text{int}^{-}(R)$] values of the propagated uncertainty,
\begin{equation}
\begin{split}
V_\text{int}^{\pm}(R) = \ 
& V_\text{BO}(R) +  V_\text{rel}(R) + V_\text{ad}(R) \\ 
& \pm \sqrt{ 
  \left( \delta V_\text{BO}(R) \right)^2 
+ \left(  \delta V_\text{rel}(R) \right)^2  
+ \left(  \delta V_\text{ad}(R) \right)^2 
} \ .
\end{split}
\end{equation}
We interpolate the functions $V_\text{int}^{\pm}(R)$ similarily to $V_\text{int}(R)$. The final value of uncertainty for a given quantity is the absolute value of the difference between the value of the quantity calculated with $V_\text{int}^{+}$ and $V_\text{int}^{-}$ divided by 2.

\subsection{Monomer energies}
\label{sec:monomer}

Determining the uncertainty of molecular electronic structure calculations is a challenging task~\cite{Chung_2016}. To verify the accuracy and reliability of our approach to estimating the CBS limit and numerical uncertainties of the interaction energy, we apply the procedure described in Sec.~\ref{sec:uncertainties} to the energies of the monomers. We report the present results and compare them with previous accurate numerical and analytical results in Table~\ref{TABLE_energies_of_monomers}.

\begin{table}[tb!]
\caption{The non-relativistic Born-Oppenheimer energy, $E_\text{BO}$, of the monomers: He($^1\!S$), He$^{*}$($^3\!S$), and He$^{+}$($^2\!S$), the adiabatic correction, $E_\text{ad}$, and various components of the relativistic correction, $E_\text{MV}$, $E_\text{D1}$, $E_\text{CG}=E_\text{MV}+E_\text{D1}$, $E_\text{D2}$, and $E_\text{OO}$, from the present atomic calculations compared with the previous, highly accurate, numerical and analytical results reported with 12 exact digits after the decimal point only. All values are in atomic units. 
\label{TABLE_energies_of_monomers}} 
\begin{ruledtabular}
\begin{tabular}{ld{2.13}d{2.16}d{2.13}}
&
\multicolumn{1}{c}{He($^1\!S$) } &
\multicolumn{1}{c}{He$^{*}$($^3\!S$)} &
\multicolumn{1}{c}{He$^{+}$($^2\!S$) } \\
\hline
$E_\text{BO}$& -2.903724(110) & -2.17522928(73) & -1.999999993319 \\
&
-2.903724377034\footnote{\label{note:Pachucki2010}Ref.~\cite{yerokhin_pachucki_2010}.}&
-2.175229378236\textsuperscript{\ref{note:Pachucki2010}}&
-2.0\footnote{\label{note:exact}Exact analytical value.} \\
\hline
$E_\text{ad}$& 
0.000419895(54)&0.00029922975(80)&0.000274186775 \\
& 0.000419888686\footnote{\label{note:Drake}Ref.~\cite{DRAKE19887}.}
& 0.000299229761\textsuperscript{\ref{note:Drake}}
& 0.000274186711\textsuperscript{\ref{note:exact}}
\\
\hline
$E_\text{MV}$&-0.00071945(52)&-0.00055667041(78) &-0.000532244084 \\
& -0.000720065710\textsuperscript{\ref{note:Drake}} 
& -0.000556949803\textsuperscript{\ref{note:Drake}} 
& -0.000532513545\textsuperscript{\ref{note:exact}} 
\\
$E_\text{D1}$& 0.000605359(50)&0.000441495752(10)&0.000425741438 \\
& 0.000605748157\textsuperscript{\ref{note:Drake}} 
& 0.000441775136\textsuperscript{\ref{note:Drake}} 
& 0.000426010836\textsuperscript{\ref{note:exact}} 
\\
$E_\text{CG}$&-0.00011409(47)&-0.00011517466(79)&-0.000106502646 \\
& -0.000114317553\textsuperscript{\ref{note:Drake}} 
& -0.000115174667\textsuperscript{\ref{note:Drake}} 
& -0.000106502709\textsuperscript{\ref{note:exact}} 
\\
$E_\text{D2}$&-0.0000178(14)&0.0&\multicolumn{1}{c}{-} \\
& -0.000017790950\textsuperscript{\ref{note:Drake}}
& 0.0\textsuperscript{\ref{note:exact}}
\\
$E_\text{OO}$&-0.000007399(89)&-0.00000008672(17)&\multicolumn{1}{c}{-} \\
& -0.000007406981\textsuperscript{\ref{note:Drake}} 
& -0.000000086716\textsuperscript{\ref{note:Drake}}
\\
\end{tabular}
\end{ruledtabular}
\end{table}

We extrapolate the BO energy of the singlet and triplet helium atom assuming the CBS-limit convergence behavior analogous to Eqs.~\eqref{extrapolation:dublet} and \eqref{extrapolation:quartet}, respectively. The calculations for the ground state singlet helium atom are performed using the d$X$Z$(^1\!S)$ family of basis sets with cardinal numbers up to $X=8$. For the metastable triplet helium atom, we use the d$X$Z$(^3\!S)$ family of basis sets with cardinal numbers up to $X=7$. In both cases the numerical uncertainty is estimated as a difference between the extrapolated value and the result calculated with the largest basis set. The present BO energies show a very good agreement with the previous highly-accurate data from Ref.~\cite{yerokhin_pachucki_2010}. Additionally, the comparison suggests that our estimation of the uncertainty is too conservative, especially in the case of the singlet helium atom. This is a strong indication that our extrapolation procedure may provide highly accurate results also in the more challenging calculations of the molecular BO interaction energy, assuring at the same time reliable estimations of error bars.

We also extrapolate the atomic post-BO corrections using the extrapolation formula analogous to Eq.~\eqref{extrapolationCorrection}. Following the approach adopted in Ref.~\cite{cencek_2012}, we determine the values of the exponents $n_Y$ by fitting the expression $A/X^{n_Y}$ to the difference between the corrections calculated using families of basis sets with increasing cardinal number $X$, $E_Y^X$, and the reference atomic values~\cite{DRAKE19887}. The adiabatic and relativistic corrections are calculated with basis sets from the d$X$Zcp and d$X$Zu series, respectively, of the $(^1\!S)$ type in the case of the singlet helium atom and the $(^3\!S)$ type for the triplet helium atom. In the fitting, we use the results obtained with basis sets with cardinal numbers $X\ge5$. We conclude the following optimal exponents for the singlet atom: $n_\text{ad}=3$, $n_\text{CG}=1.5$, $n_\text{D2}=1$, and $n_\text{OO}=1.5$ for the adiabatic, Cowan-Griffin, two-electron Darwin, and orbit-orbit correction, respectively. The corresponding exponents for the triplet atom are: $n_\text{ad}=5$, $n_\text{CG}=3.5$, and $n_\text{OO}=3.5$. For the one-electron relativistic corrections we take the same exponents as for the Cowan-Griffin correction, that is, $n_\text{MV}=n_\text{D1}=n_\text{CG}$. In the extrapolations of the post-BO corrections to the molecular interaction energy, we assume that the same exponents as for the singlet and triplet helium atom are applicable for the doublet and quartet molecular states, respectively. Note, that we do not need to extrapolate the two-electron Darwin correction for the triplet helium atom and quartet molecular states. For these spin-polarized states, the electronic wavefunction is antisymmetric with respect to the interchange of spatial coordinates of any pair of electrons, so the wavefunction is equal to zero when the electrons approach each other and the two-electron Darwin correction, defined by Eq.~\eqref{d2}, vanishes identically.

In Table~\ref{TABLE_energies_of_monomers} we present the results obtained by applying our extrapolation procedure, defined by the formula Eq.~\eqref{extrapolationCorrection} with the exponents $n_Y$ discussed above, for atomic post-BO corrections. In the extrapolations, we use the values of a given correction calculated with the two largest cardinal numbers $X$ available in the appropriate basis set family. The uncertainty is then estimated as the difference between the extrapolant an the result calculated using basis set with the largest $X$. In the case of the adiabatic, Cowan-Griffin, and two-electron relativistic corrections, the previous accurate numerical results from Ref.~\cite{DRAKE19887} are always within our estimated error bars. By contrast, the uncertainties of the mass-velocity and one-electron Darwin corrections are underestimated, even by four orders of magnitude in the case of $E_\text{D1}$ for the triplet helium atom. This observation supports our decision to extrapolate the sum of the one-electron relativistic corrections---the Cowan-Griffin correction---rather than each of them separately.

As the helium ion is a one-electron system where the electron-correlation effects are absent, its energy and energy corrections exhibit very fast (exponential) convergence to the CBS limit with the basis set's cardinal number~\cite{HALKIER1999437}. In Table~\ref{TABLE_energies_of_monomers}, we provide the results for the helium ion obtained with large basis sets d7Z$(^3\!S)$ (BO energy), d7Zcp$(^3\!S)$ (adiabatic correction), and d7Zu$(^3\!S)$ (one-electron relativistic corrections), that can be considered as converged. We see that $E_\text{BO}$, $E_\text{ad}$, and $E_\text{CG}$ agree well with the analytical results, the differences being of the order of $10^{-9}$~$E_\text{h}$ or less. Again, the good agreement of the Cowan-Griffin correction is a result of error cancellation, as the errors of the individual one-electron relativistic correction are three orders of magnitude larger.

\subsection{Long-range interaction potential}

At large internuclear distances, where the magnitude of the interaction energy becomes smaller and smaller, the relative accuracy of the super-molecule approach decreases as more and more significant digits cancel out when the energies of atom and ion are subtracted from the dimer energy. However, the interaction energy in this range may be crucial for the ultracold collisional properties of the system. Thus, we connect the calculated short- and intermediate-range PECs with the long-range form of the interaction energy, which for the $S$-state atom and the $S$-state ion is given by
\begin{equation}
V_\text{int}(R) = -\sum_{n=4}^\infty\frac{C_n}{R^n}  \ ,
\label{LRF}
\end{equation}
where the $n=5$ term is absent and $C_n$ are the asymptotic coefficients. To assure qualitatively correct description of the long-range part of the interaction energy, we can restrict ourselves only to the asymptotic expansion of its dominant, BO part. This is sufficient, as the post-BO corrections are about four orders of magnitude smaller---the adiabatic correction being of the order of the electron-to-nucleus mass ratio and the relativistic correction of the order $\alpha^2$. What is more important, unlike in the case of the interaction between two neutral $S$-state atoms, the post-BO corrections in the atom-ion system do not introduce terms vanishing like powers of $1/R$ other than that already present in the expansion of the BO energy~\cite{PRZYBYTEK2012170,Meath1966}.

In our representation of the long-range expansion of the BO interaction energy, we include all effects from the polarization perturbation theory that give raise to terms vanishing with the distance as $1/R^9$ or slower. The $C_n$ coefficients with $n\le9$ can then be represented as a sum of three parts of different physical origin,
\begin{equation}
    C_n = C_n^\text{ind} + C_n^\text{disp} +C_n^\text{ind-disp} \ ,
\end{equation}
where the $C_n^\text{ind}/R^n$ terms describe up to fourth-order induction effects coming from the electrostatic interaction between the charge of the ion and the induced electric multipole moments of the atom, the $C_n^\text{disp}/R^n$ terms describe the second-order dispersive interaction between instantaneous multipole-induced multipole moments of the ion and the atom stemming from quantum fluctuations, and the $C_n^\text{ind-disp}/R^n$ terms describe the third-order interaction of a mixed type. Specifically, we obtain the induction coefficients using properties of the monomers as: $C_4^\text{ind} = \frac{q^2 \alpha_1}{2}$, $C_6^\text{ind} = \frac{q^2 \alpha_2}{2}$, $C_7^\text{ind} = -\frac{q^3 \beta_{112}}{2}$, $C_8^\text{ind} = \frac{q^2 \alpha_3}{2} + \frac{q^4 \gamma_{1111}}{24}$, and $C_9^\text{ind} = -q^3\beta_{123} - \frac{q^3\beta_{222}}{6}$, where $q = 1$ is the charge of the helium ion, and $\alpha_l$ are the $2^l$-pole polarizabilities, $\beta_{l_1 l_2 l_3}$ are the first hyperpolarizabilites, and $\gamma_{l_1 l_2 l_3 l_4}$ are the second hyperpolarizabilities of the helium atom. The $C_4$ and $C_7$ asymptotic coefficients comprise only the induction part. The only nonzero dispersion coefficients are $C_6^{\text{disp}}$ and $C_8^{\text{disp}}$, and the only nonzero induction-dispersion coefficient is $C_9^\text{ind-disp}$.

\begin{table}[t!]
\caption{The static $2^l$-pole polarizabilities ($\alpha_l$) and the first ($\beta_{l_1 l_2 l_3}$) and second ($\gamma_{l_1 l_2 l_3 l_4}$) hyperpolarizabilities of the He atom in the $^1\!S$ and $^3\!S$ state. The second-order dispersion ($C_6^\text{disp}$, $C_8^\text{disp}$) and third-order induction-dispersion ($C_9^\text{ind-disp}$) asymptotic coefficients for the He atom in the $^1\!S$ and $^3\!S$ states interacting with the ground-state He$^+$ ion. All values are in atomic units.}
\label{tab:C_ns}
\begin{ruledtabular}
\begin{tabular}{ld{3.12}d{5.8}}
& 
\multicolumn{1}{c}{$\mathrm{He}(^1\!S)$} &
\multicolumn{1}{c}{$\mathrm{He}^*(^3\!S)$} \\
\hline
$\alpha_1$       & 1.3832 & 315.619 \\
                 & 1.38319217440\footnote{\label{note:Yan:96}Ref.~\cite{Yan:96}.}
                 & 315.631468\footnote{\label{note:Yan:98}Ref.~\cite{Yan:98}.} \\
$\alpha_2$       & 2.4451 & 2707.48  \\
                 & 2.445083101\textsuperscript{\ref{note:Yan:96}}
                 & 2707.8773\textsuperscript{\ref{note:Yan:98}}  \\
$\alpha_3$       & 10.6138 & 88271.9  \\
                 & 10.6203286\textsuperscript{\ref{note:Yan:96}}
                 & 88377.3253\textsuperscript{\ref{note:Yan:98}}  \\
$\beta_{112}$    & -7.327  & -1.601\times 10^5 \\
                 & -7.327\footnote{\label{note:Bishop89}Ref.~\cite{Bishop:89}.} 
                 & \\
$\beta_{123}$    & -30.47 & -2.943\times 10^6 \\
$\beta_{222}$    & -17.94 & -1.316\times 10^6 \\
$\gamma_{1111}$  & 42.98 & -5.790\times 10^6 \\
                 & 43.10\textsuperscript{\ref{note:Bishop89}} 
                 & \\
\hline
&
\multicolumn{1}{c}{$\mathrm{He}(^1\!S)+\mathrm{He}^+(^2\!S)$} &
\multicolumn{1}{c}{$\mathrm{He}^*(^3\!S)+\mathrm{He}^+(^2\!S)$} \\
\hline
$C^\text{disp}_6$       & 0.3745 & 6.079 \\
                        & 0.3743\footnote{\label{note:Davison66}Ref.~\cite{Davison:66}.}
                        & \\
$C^\text{disp}_8$       & 2.714 & 348.2 \\
                        & 2.712\textsuperscript{\ref{note:Davison66}}
                        & \\
$C^\text{ind-disp}_9$   & 4.382 & 11450 \\
\end{tabular}
\end{ruledtabular}
\end{table}

Table~\ref{tab:C_ns} presents calculated values of the $2^l$-pole polarizabilities and the first and second hyperpolarizabilities of He in the $^1\!S$ and $^3\!S$ states, as well as the calculated values of the dispersion and induction-dispersion coefficients. The calculations are performed using the sum-over-states formulas involving electric-multipole transition moment matrix elements between the electronic ground and excited states of the monomers described at the FCI level of theory. In the calculations of the atomic (hyper)polarizabilities, we use the d$X$Z$(^1\!S)$ and d$X$Z$(^3\!S)$ family of basis sets for the singlet and helium atom, respectively. In the calculations of the dispersion and induction-dispersion coefficients, the basis set for the helium ion is the same as the basis set for the helium atom the ion interacts with. The present values agree well with available previous results~\cite{Yan:96,Yan:98,Bishop:89,Davison:66}. The very accurate $2^l$-pole polarizabilities of the singlet and triplet helium atom taken from Refs.~\cite{Yan:96} and \cite{Yan:98}, respectively, are used in the final calculation of the induction coefficients $C_n^\text{ind}$.

We connect the spline-interpolated potentials from Eq.~\eqref{eq:Vint} and the long-range functions from Eq.~\eqref{LRF} at $R=47.5$~a$_0$, where the differences between them are numerically negligible for all electronic states. We also confirmed the convergence of the scattering parameters (i.e., the scattering lengths) with regard to the position of the connection.

\subsection{Nuclear dynamics calculations}

The collisional properties of the system are described by the radial part, $\chi(R)$, of the total wavefunction within the partial-wave decomposition. We obtain $\chi(R)$ by solving the nuclear Shr\"odinger equation in the adiabatic approximation 
\begin{equation}
\left[ 
- \frac{\hbar^2}{2 \mu}\frac{d^2}{dR^2} 
+ \frac{\hbar^2 J(J+1)}{2 \mu R^2} 
+ V_\text{int}(R)
\right] \chi(R)= E_\text{nu} \, \chi(R) \ ,
\label{nuclear_Shro_2}
\end{equation}  
where $J$ is the rotational quantum number and $\mu$ is the reduced mass of the system calculated using the atomic, rather than nuclear, masses of the helium isotopes to account for the leading, nonadiabatic effects. For $E_\text{nu} > 0$ the function $\chi(R)$ represents the scattering state describing the motion of nuclei on the electronic interaction potential $V_\text{int}(R)$. In this case $E_\text{nu}$ is the relative collision energy of the ion-atom system (denoted $E_\text{col}$). For $E_\text{nu} < 0$ the function $\chi(R)$ becomes the radial wavefunction of the bound rovibrational state of the molecule and $E_\text{nu}$ becomes the binding energy, denoted $E_{\nu,J}$, where $\nu$ is the vibrational quantum number.

To find the scattering wavefunction, we solve Eq.~\eqref{nuclear_Shro_2} using the renormalized Numerov propagator~\cite{johnson_1978}. We impose the long-range scattering boundary conditions on $\chi(R)$ in terms of the Bessel and von Neumann functions. We then calculate the $K$ reactance matrix which gives the scattering $S$ matrix. We evaluate the scattering length $a$ from the $S$ matrix which in the $s$-wave regime ($J=0$) is given by
\begin{equation}
a = \frac{1}{i k} \frac{1-S_{00}}{1+S_{00}} \ ,
\end{equation}
where $k = \sqrt{2\mu E_\text{col} }/\hbar$ is the wavenumber of the scattering state and $S_{00}$ is a diagonal matrix element of the $S$-matrix. We set the collision energy $E_\text{col}$ at $10^{-10}$~K.

To find the bound eigenstates, we employ the discrete variable representation (DVR) method~\cite{tiesinga_williams_julienne_1998} based on finding eigenvalues of the shifted inverse or Green-function operator. We calculate the bound-state nuclear wavefunctions and the binding energies for a given molecular state potential $V_\text{int}(R)$ which are parametrized by the quantum numbers $\nu$ and $J$. We use a logarithmic grid function and a grid of 3000 values of $R$ ranging from $R_0 = 0.5$~a$_0$ to $R_f = 10^4$~a$_0$.

The control of collisions can be realized by tuning the interaction strength using Feshbach resonances once the $s$-wave scattering regime is reached~\cite{RevModPhys.82.1225}. For the ion-atom systems, the collisional energy needed to reach the $s$-wave regime is at least two orders of magnitude smaller than in neutral systems due to lower values of the characteristic energy of interaction, $E_4 = \frac{\hbar^2 }{2 \mu R_4^2}$, where $R_4 = \sqrt{\frac{2 \mu  C_4}{\hbar^2}}$ is the characteristic length of interaction~\cite{TomzaRMP19}. $R_4$ typically establishes the order of magnitude of the scattering length. Light systems, like He$_2^+$, have higher values of $E_4$ and  are therefore promising for easier reaching the $s$-wave regime.

\subsection{Spectroscopic parameters}

We perform a detailed analysis of the rovibrational structure of the He$_2^+$ molecular ion. We calculate the molecular spectroscopic constants by fitting the following expansion to the obtained rovibrational energies
\begin{equation}
\label{Dunham}
\begin{split}
E_{\nu,J}  = \ 
&   \omega_e \left(\nu+\frac12\right) - \omega_e x_e \left(\nu+\frac12\right)^2 \\   
& + \omega_e y_e \left(\nu+\frac12\right)^3 + \omega_e z_e \left(\nu+\frac12\right)^4 \\  
& + B_e \left[J(J+1)\right] - \alpha_e \left(\nu+\frac12\right)\left[J(J+1)\right] \\ 
& + \gamma_e \left(\nu+\frac12\right)^2 \left[J(J+1)\right] \\
& - \mathcal{D}_e \left[J(J+1)\right]^2 + \beta_e \left(\nu+\frac12\right)\left[J(J+1)\right]^2 \\
& + H_e \left[J(J+1)\right]^3 - D_e \ ,
\end{split}
\end{equation}
where $\nu$ and $J$ are the vibrational and rotational quantum numbers, respectively, $\omega_e$ is the harmonic constant, $ \omega_e x_e$, $\omega_e y_e$, and $\omega_e z_e$ are the anharmonicity constants. We include the dependence of the rotational constant $B_\nu$ on the vibrational quantum number
\begin{equation}
B_\nu=B_e -\alpha_e \left(\nu+\frac12\right) + \gamma_e\left(\nu+\frac12\right)^2 \ ,
\end{equation}
where $B_e$ is the rotational constant at the equilibrium distance, and $\alpha_e$ and $\gamma_e$ are the first- and second-order vibration-rotation interaction constants, respectively.  We also include higher-order rotational constants and their vibrational dependence, i.e., the first- and second-order centrifugal distortion constants $\mathcal{D}_e$ and $H_e$, and the centrifugal-vibration interaction constant $\beta_e$.

The change of the equilibrium rotational constants caused by nonadiabatic effects $\Delta B_e$ is estimated by~\cite{doi:10.1021/cr60292a003}
\begin{equation}
\label{eq:NAE}
\Delta B_e= \frac{m_e}{m_p} g B_e \ ,
\end{equation}
where $\frac{m_e}{m_p}$ is the electron-to-proton mass ratio, and $g$ is the rotational $g$-factor, which is computed using the natural connection and London orbitals~\cite{doi:10.1063/1.472143} implemented in the \textsc{Dalton} 2020 program~\cite{Dalton2020}. The computations of $g$ employ electronic wavefunction obtained with the FCI method and doubly augmented correlation consistent quadruple-zeta basis set d-aug-cc-pVQZ~\cite{woon1994a}.

\section{Numerical results and discussion}
\label{sec:results}
\subsection{Potential energy curves}

\begin{figure}[t!]
\includegraphics[width=\columnwidth]{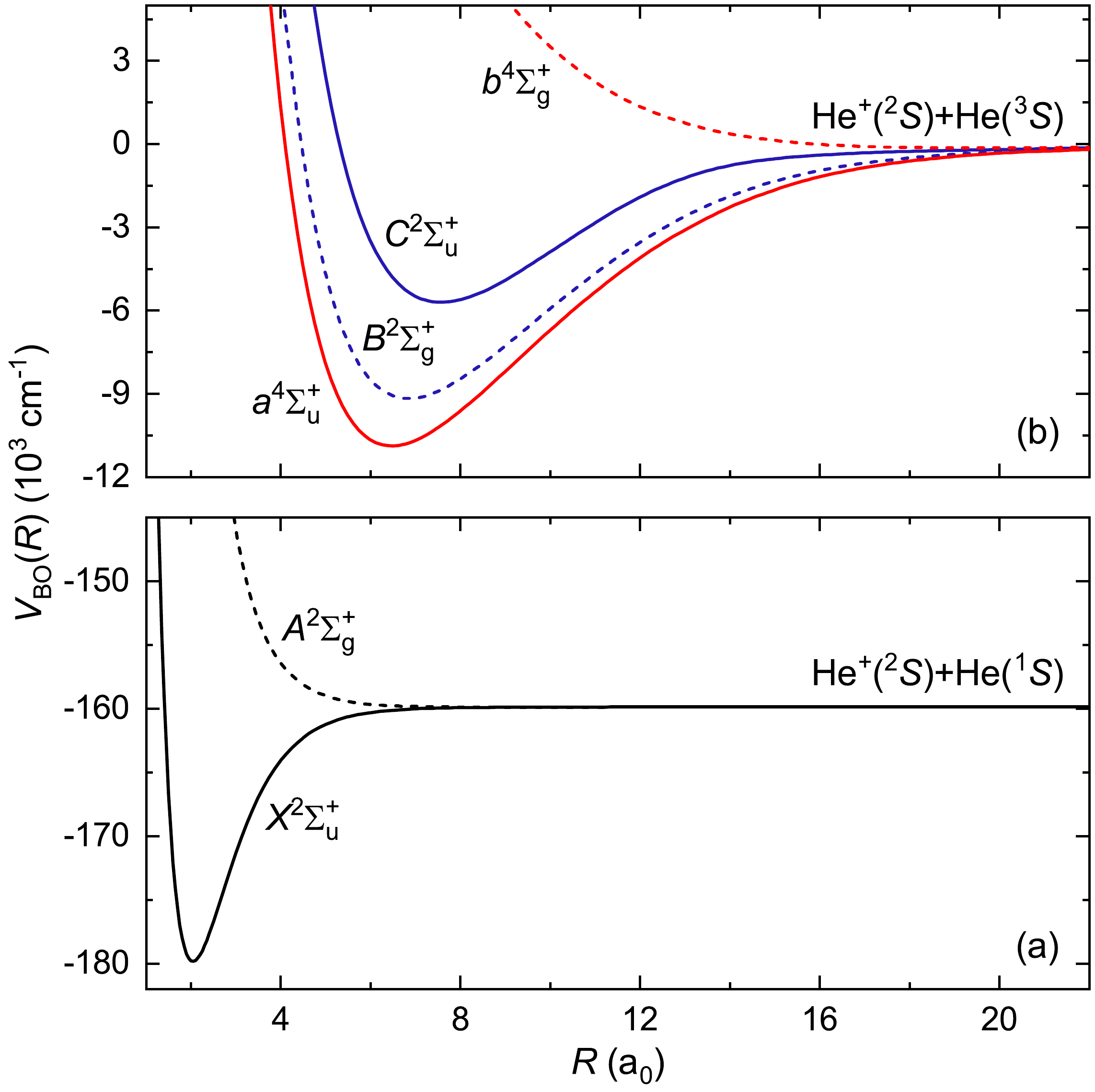}
\caption{The non-relativistic Born-Oppenheimer potential energy curves of the He$_2^+$ molecular ion in the (a) ground and first-excited and (b) higher-excited electronic states.}
\label{fig:PES_ext_no_corr}
\end{figure}

Figure~\ref{fig:PES_ext_no_corr} presents the calculated PECs for the \stX, \stA, \sta, \stB, \stC, and \stb{} molecular electronic states of the He$_2^+$ molecular ion listed in the order of increasing energy. We set the asymptotic energy to zero for the higher-excited (\sta -- \stb) molecular states resulting from the He$^+$($^2\!S$)+He$^*$($^3\!S$) combination of interest. The positions of the minima on PECs for the higher-excited electronic states are at larger distances as compared with the position of the minimum for the ground state. This results from a larger van der Waals radius and larger atomic polarizability of the helium atom in the triplet excited state as compared with the properties of the helium atom in the singlet ground state. The three higher-excited electronic states (\sta, \stA, and \stB) are relatively deeply-bound but half as deep as the ground \stX{} state. The \stb{} state is weakly-bound, however, deeper than the first-excited \stA{} state, which is very weakly bound.

\begin{figure}[!t]
\includegraphics[width=\columnwidth]{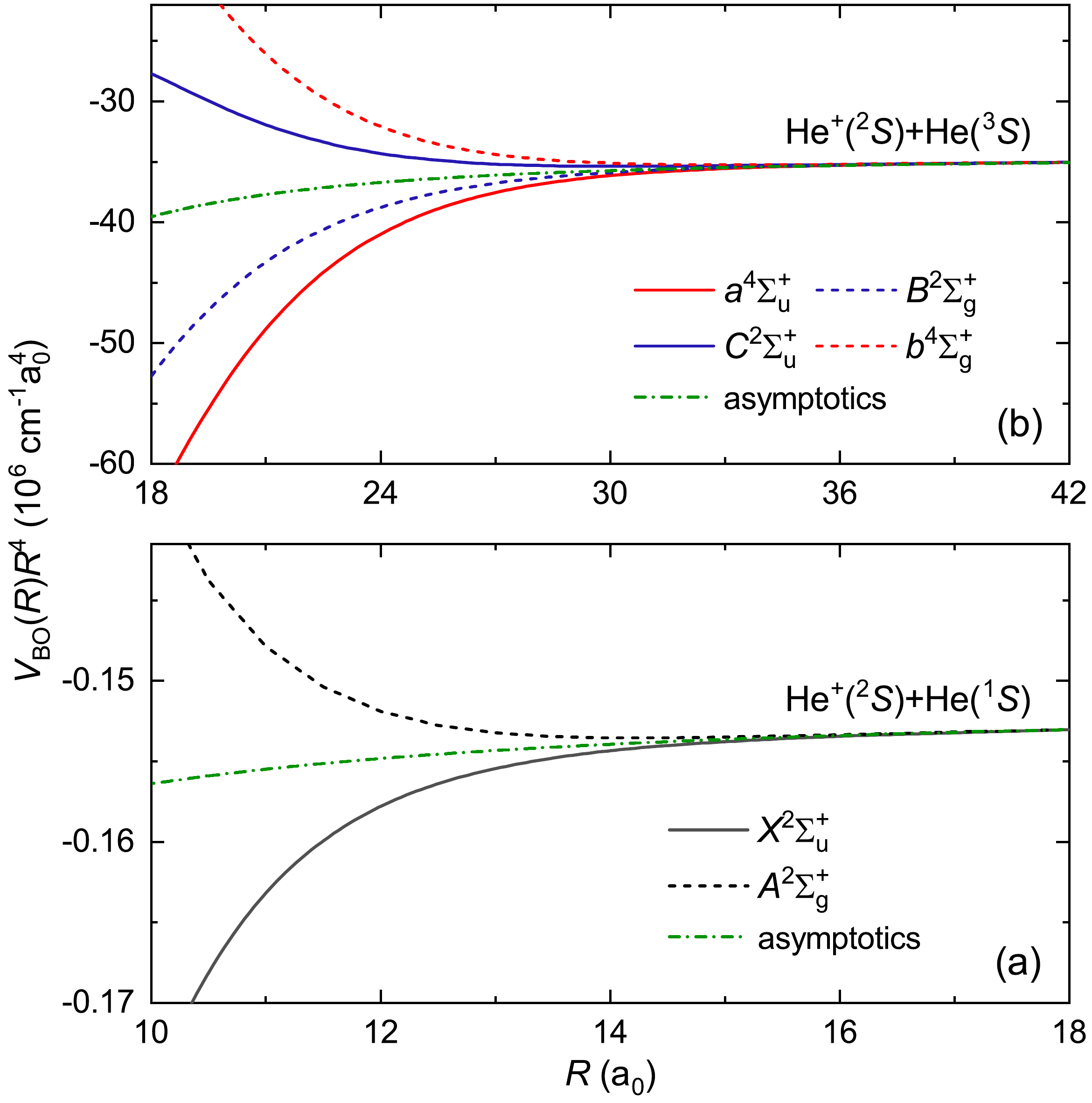}
\caption{The asymptotics of the potential energy curves of the He$_2^+$ molecular ion. The scaled interaction energy of the (a) ground and first-excited and (b) higher-excited molecular electronic states. Note the different scales in panels (a) and (b).}
\label{fig:PEC}
\end{figure}

The long-range behavior of the calculated PECs is analyzed in Fig.~\ref{fig:PEC}, where we present the interaction energies, multiplied by $R^4$, at large internuclear distances. The curves labeled as 'asymptotics' are given by the long-range expansion from Eq.~\eqref{LRF}, also multiplied by $R^4$, and differ from the constant line, specified by the $C_4$ coefficients, due to the presence of higher-order polarization terms. Figure~\ref{fig:PEC} shows a good agreement of the molecular calculations at large distances with the long-range expansion of the interaction potentials obtained independently from atomic data. At the intermediate range, the calculated curves diverge from the asymptotic expansions in a systematic way given by the electronic exchange interaction, which is the largest for the quartet states. The long-range interactions are much stronger in the higher-excited electronic states than in the ground and first-excited states because the polarizability of the helium atom in the triplet state is two orders of magnitude larger than in the singlet state. For the same reason, the higher-order polarization terms and exchange effects are important at larger distances in the excited asymptote, He$^+(^2\!S)$+He$^*(^3\!S)$, as compared with the ground one, He$^+(^2\!S)$+He$(^1\!S)$. 

\begin{table*}[!hb]
\caption{The adiabatic $V_\text{ad}$ and relativistic $V_\text{rel}$ corrections and the mass-velocity $V_\text{MV}$, one-electron Darwin $V_\text{D1}$, Cowan-Griffin $V_\text{CG}$, two-electron Darwin $V_\text{D2}$, and orbit-orbit $V_\text{OO}$ components of the relativistic correction (in $\text{cm}^{-1}$) at the respective equilibrium distances $R_e$ for the He$_2^+$ molecular ion in the ground and excited electronic states.} 
\begin{ruledtabular}
\begin{tabular}{ld{2.9}d{2.8}d{2.8}d{2.8}d{2.8}d{2.9}d{2.9}}
State&\multicolumn{1}{c}{ $V_\text{ad}(R_e)$ } & \multicolumn{1}{c}{ $V_\text{rel}(R_e)$ }  &\multicolumn{1}{c}{ $V_\text{MV}(R_e)$ }  & \multicolumn{1}{c}{$V_\text{D1}(R_e)$} &  \multicolumn{1}{c}{$V_\text{CG}(R_e)$} &\multicolumn{1}{c}{$V_\text{D2}(R_e)$} &\multicolumn{1}{c}{ $V_\text{OO}(R_e)$} \\
\hline
\stX&-2.129(2)& -0.085(29) &-3.983(4) & 3.691(10) &-0.291(6) &  -0.41(3) & 0.615(4)\\
\stA& -0.0022563(7)&0.002270(9) &0.00632(3) & -0.00482(2)&0.001491(7) & 0.000218(5) & 0.0005616(15)\\
\sta&-0.541(2)&0.4039(10) &0.9174(5) & -0.7823(13)& 0.1351(9)& 0 & 0.2689(4)\\
\stB& -0.071(3)&0.423(7) &1.1191(11) & -0.766(3)&0.354(4) & -0.065(5) & 0.1353(4)\\
\stC& 0.593(3)&0.553(16) &1.710(7) & -0.872(2)& 0.839(9)& -0.193(14) & -0.0924(8)\\
\stb& -0.05045(2)& 0.03740(2) &0.101192(8) & -0.07116(3)& 0.03004(2)&  0 & 0.007367(6)\\
\end{tabular}
\end{ruledtabular}
\label{contributions_of_rel_corrections_at_Re}
\end{table*}

\begin{figure}[tb!]	
\begin{center}
\includegraphics[width=\columnwidth]{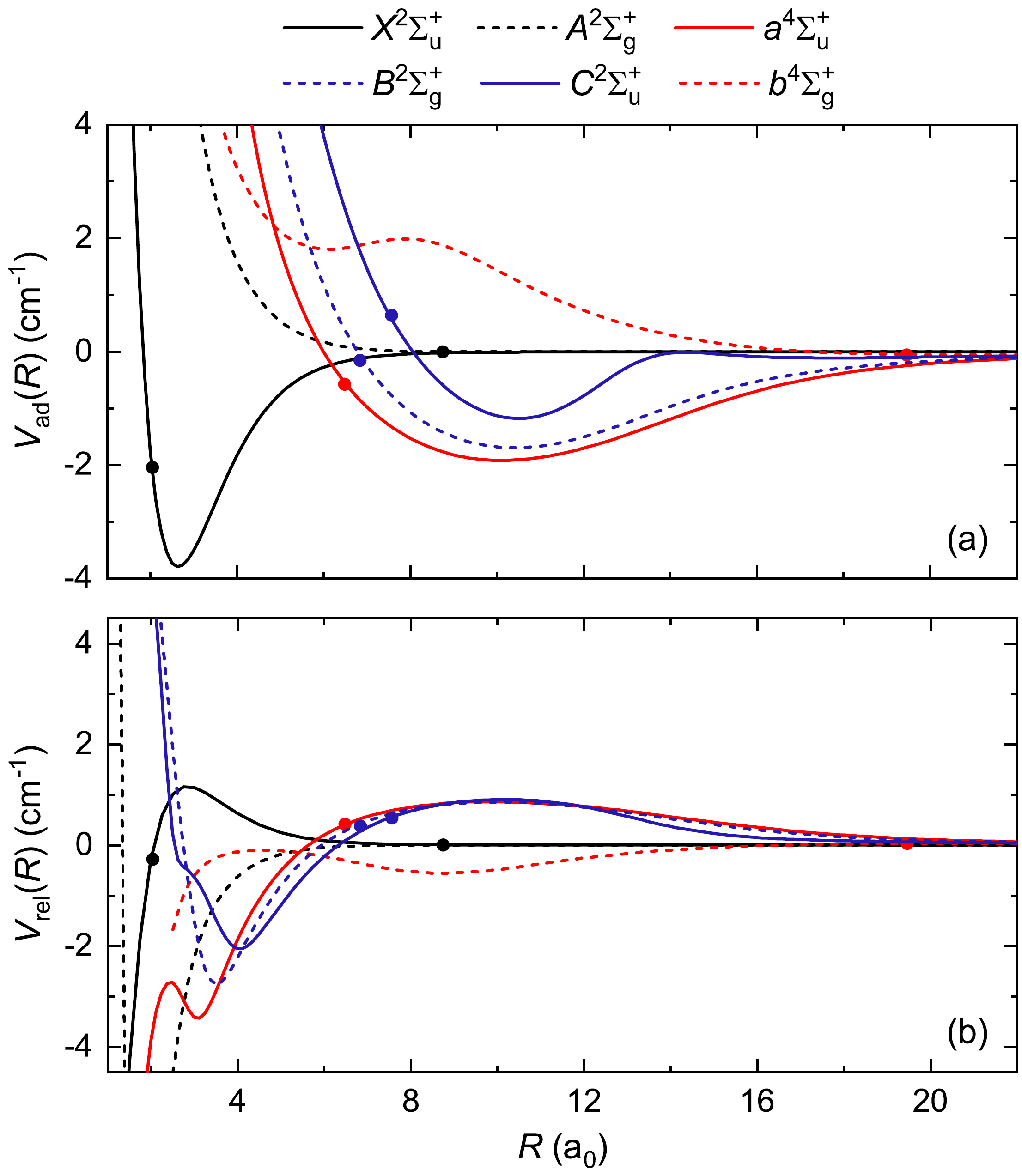}
\end{center}
\caption{The (a) adiabatic and (b) relativistic corrections to the interaction energy of the He$_2^+$ molecular ion in the ground and excited electronic states. The points indicate values for equilibrium distances of the corresponding potential energy curves.}
\label{fig:corr}
\end{figure}

\begin{figure}[tb!]
\begin{center}
\includegraphics[width=\columnwidth]{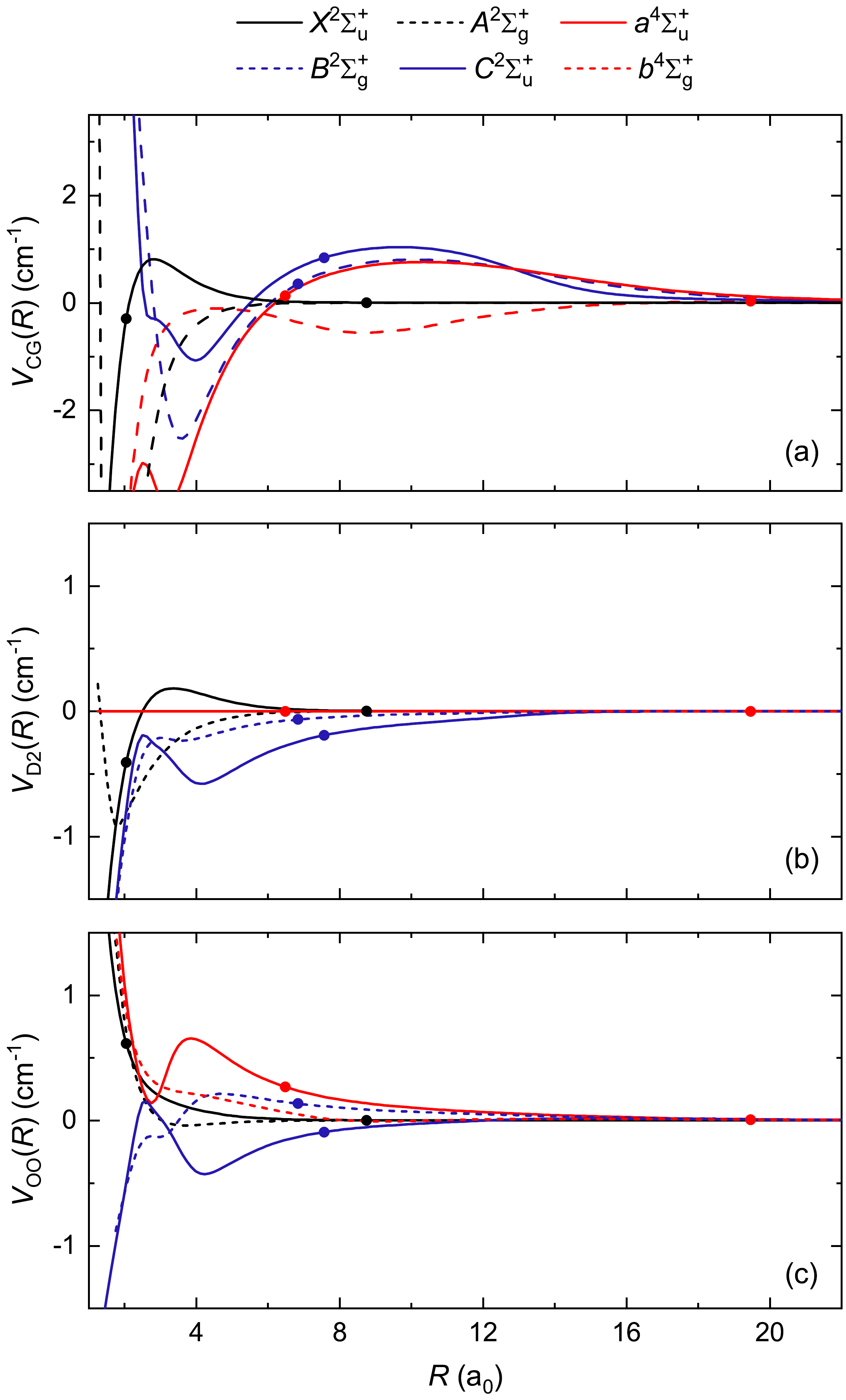}
\end{center}
\caption{The (a) Cowan-Griffin, (b) two-electron Darwin, and (c) orbit-orbit components of the relativistic correction for the He$_2^+$ molecular ion in the ground and excited electronic states. The points indicate values for equilibrium distances of the corresponding potential energy curves.}
\label{fig:rels_all_horizontally}
\end{figure}

Figure~\ref{fig:corr} presents the calculated adiabatic and relativistic corrections to the PECs. Both corrections have values of the same order of magnitude, but they have opposite signs and partially cancel out. Thus, the final uncertainties of our PECs may be slightly overestimated due to the cancellation of errors after including both corrections. Additionally, Figure~\ref{fig:rels_all_horizontally} shows the calculated Cowan-Griffin, two-electron Darwin, and orbit-orbit components of the relativistic corrections. As discussed in Sec.~\ref{sec:monomer}, the $V_\text{D2}(R)$ terms are indentically equal to zero for the quartet states. Table~\ref{contributions_of_rel_corrections_at_Re} presents the values of the adiabatic and relativistic corrections and different components of the relativistic correction at respective equilibrium distances, $R_e$, of different molecular electronic states. The mass-velocity and one-electron Darwin terms have opposite signs and partially cancel out resulting in the Cowan-Griffin correction of the order of the two-electron Darwin and orbit-orbit components.  The ground \stX{} state has the largest value of the adiabatic correction while three deeply-bound higher-excited states have the largest values of the relativistic contribution at $R_e$ among the studied molecular electronic states. Our values of the relativistic corrections to the total energy of the \stX{} state near its equilibrium distance agree well, up to around 0.1$\,\text{cm}^{-1}$,  with previous, accurate results from Ref.~\cite{FerencPRL20}.

Both adiabatic and relativistic corrections have values of the order of the uncertainties of our non-relativistic Born-Oppenheimer PECs. Their inclusion allows, however, to evaluate their impact on vibrational and scattering properties and estimate the theory improvement needed to match the accuracy of possible spectroscopic measurements and scattering experiments.

We estimate the quantum electrodynamics corrections to the interaction energy and get $V_{\text{QED},1}(R)/V_\text{D1}(R) = 0.038 $ for He$^+(^2\!S)$+He$(^1\!S)$, and $V_{\text{QED},1}(R)/V_\text{D1}(R) = 0.0378 $ and $ V_{\text{QED},2}(R)/V_\text{D2}(R) = 0.0396$  for He$^+(^2\!S)$+He$^*(^3\!S)$. Including all post-Breit-Pauli corrections is therefore unnecessary at the intended level of accuracy, because their contribution to the interaction energy is a small fraction of the relativistic correction and final PEC uncertainty.

\begin{table*}[!tbh]
\caption{Spectroscopic characteristics of the He$_2^+$ molecular ion in the ground and excited electronic states: equilibrium bond lengths $R_e$, well depths $D_e$, harmonic constants $\omega_e$, and  dissociation energies $D_0$ for the $^4$He$_2^+$ isotopologue. \label{tab:spec1}} 
\begin{ruledtabular}
\begin{tabular}{lld{9}d{9}d{9}d{9}}
State & Potential &
\multicolumn{1}{c}{ $R_e\,$($\text{a}_0$) } &
\multicolumn{1}{c}{ $D_e\,$(cm$^{-1}$) } &
\multicolumn{1}{c}{ $\omega_e\,$(cm$^{-1}$) } &
\multicolumn{1}{c}{ $D_0\,$(cm$^{-1}$) } \\
\hline
\multicolumn{6}{c}{He$^+$(${}^2\!S$)+He(${}^1\!S$)} \\
\stX
& $V_\text{BO}$                              & 2.0421(1)&19954.8(5.0)&1696.8(2)&19114.2(4.9) \\
& $V_\text{BO}$+$V_\text{rel}$	             & 2.0411(1)&19954.8(5.0)&1696.8(2)&19114.2(4.9) \\
& $V_\text{BO}$+$V_\text{rel}$+$V_\text{ad}$ & 2.0421(1)&19957.0(5.0)&1696.9(2)&19116.4(4.9) \\
& \ \ theor.\footnote{Spectroscopic constants calculated using the accurate PEC from Ref.~\cite{TungJCP12}, which included only the adiabatic corrections.}
                                             & 2.0422   &19956.7     &1695.28  & 19116.2 \\
\stA
&$V_\text{BO}$                              & 8.7198(7)&17.367(5)&23.16(1)&8.077(3) \\
&$V_\text{BO}$+$V_\text{rel}$               & 8.7197(7)&17.364(5)&23.16(1)&8.075(3) \\
&$V_\text{BO}$+$V_\text{rel}$+$V_\text{ad}$ & 8.7204(7)&17.367(5)&23.16(1)&8.077(3) \\
& \ \ theor.\footnote{Spectroscopic constants taken from Ref.~\cite{xie2005}.}
                                            & 8.749    &17.382  & -       & - \\
& \ \ theor.\footnote{\label{note:Augusto2013}Spectroscopic constants taken from Ref.~\cite{augustovicova2013}.}
                                            & 8.718    &17.516  & -       &8.185 \\
\hline
\multicolumn{6}{c}{He$^+$(${}^2\!S$)+He$^*$(${}^3\!S$)} \\
\sta
&$V_\text{BO}$	                            & 6.4699(3)&10873.7(3.0)&315.8(1)&10716.3(3.1) \\
&$V_\text{BO}$+$V_\text{rel}$               & 6.4698(3)&10873.3(3.0)&315.8(1)&10715.9(3.1) \\
&$V_\text{BO}$+$V_\text{rel}$+$V_\text{ad}$	& 6.4704(3)&10873.9(3.0)&315.7(1)&10716.5(3.1) \\
\stB
&$V_\text{BO}$                              & 6.8367(7)&9167.5(6.6)&295.5(3)&9020.3(6.9) \\
&$V_\text{BO}$+$V_\text{rel}$               & 6.8365(7)&9167.0(6.6)&295.5(3)&9019.9(6.9) \\
&$V_\text{BO}$+$V_\text{rel}$+$V_\text{ad}$ & 6.8373(7)&9167.1(6.6)&295.5(3)&9020.0(6.9) \\
& \ \ theor.\textsuperscript{\ref{note:Augusto2013}.} & 6.837 & 9172.316 & - & 9024.9 \\
\stC
&$V_\text{BO}$                              & 7.565(3)&5696.8(9.8)&259.1(2.2)&\multicolumn{1}{D{.}{ }{9}}{5570(.10)} \\
&$V_\text{BO}$+$V_\text{rel}$	            & 7.565(3)&5696.2(9.8)&259.1(2.2)&\multicolumn{1}{D{.}{ }{9}}{5570(.10)} \\
&$V_\text{BO}$+$V_\text{rel}$+$V_\text{ad}$	& 7.566(3)&5695.7(9.9)&259.1(2.2)&\multicolumn{1}{D{.}{ }{9}}{5570(.10)} \\
& \ \ theor.\footnote{Spectroscopic constants taken from Ref.~\cite{augustovicova2015}.} & 7.559  & 5699.4  & - & - \\
\stb
&$V_\text{BO}$                              & 19.4137(2)&143.72(1)&25.588(2)&131.32(1) \\
&$V_\text{BO}$+$V_\text{rel}$               & 19.4135(2)&143.68(1)&25.586(2)&131.28(1) \\
&$V_\text{BO}$+$V_\text{rel}$+$V_\text{ad}$ & 19.4146(2)&143.73(1)&25.588(2)&131.28(1) \\
\end{tabular}
\end{ruledtabular}
\end{table*}

Finally, Table~\ref{tab:spec1} presents the equilibrium bond lengths $R_e$, the potential well depths $D_e$, and the harmonic constants $\omega_e$ of the calculated PECs at different levels of theory for the $^4$He$_2^+$ molecular ion in the ground and excited electronic states. We also report the dissociation energies $D_0$ of the lowest rovibrational states $v=0,J=0$.  Our calculations for the ground \stX{} state are in good agreement with previous, accurate numerical results from Ref.~\cite{TungJCP12}. This may suggest that we have achieved similar accuracy for our higher-excited states, where uncertainties are in the same order of magnitude as for the ground state. Our molecular BO energies $E_\text{BO}^{\text{He}\text{He}^+}\!(R)$  of the \stX{} electronic state also agree well, up to around 0.1$\,\text{cm}^{-1}$, with previous, accurate results from Ref.~\cite{FerencPRL20}. Finally, the parameters of our PECs agree well with other less accurate calculations for the \stX, \stA, \stB, and \stC{} electronic states~\cite{ ACKERMANN199175,carrington1995,xie2005,augustovicova2013,augustovicova2015, viehland2016}. The overall agreement of our results with previous data and the minimal effect of the adiabatic and relativistic corrections on spectroscopic constants may indicate that we have conservatively estimated (overestimated) the uncertainties of our calculations.

\begin{figure}[b!]
\begin{center}
\includegraphics[width=\columnwidth]{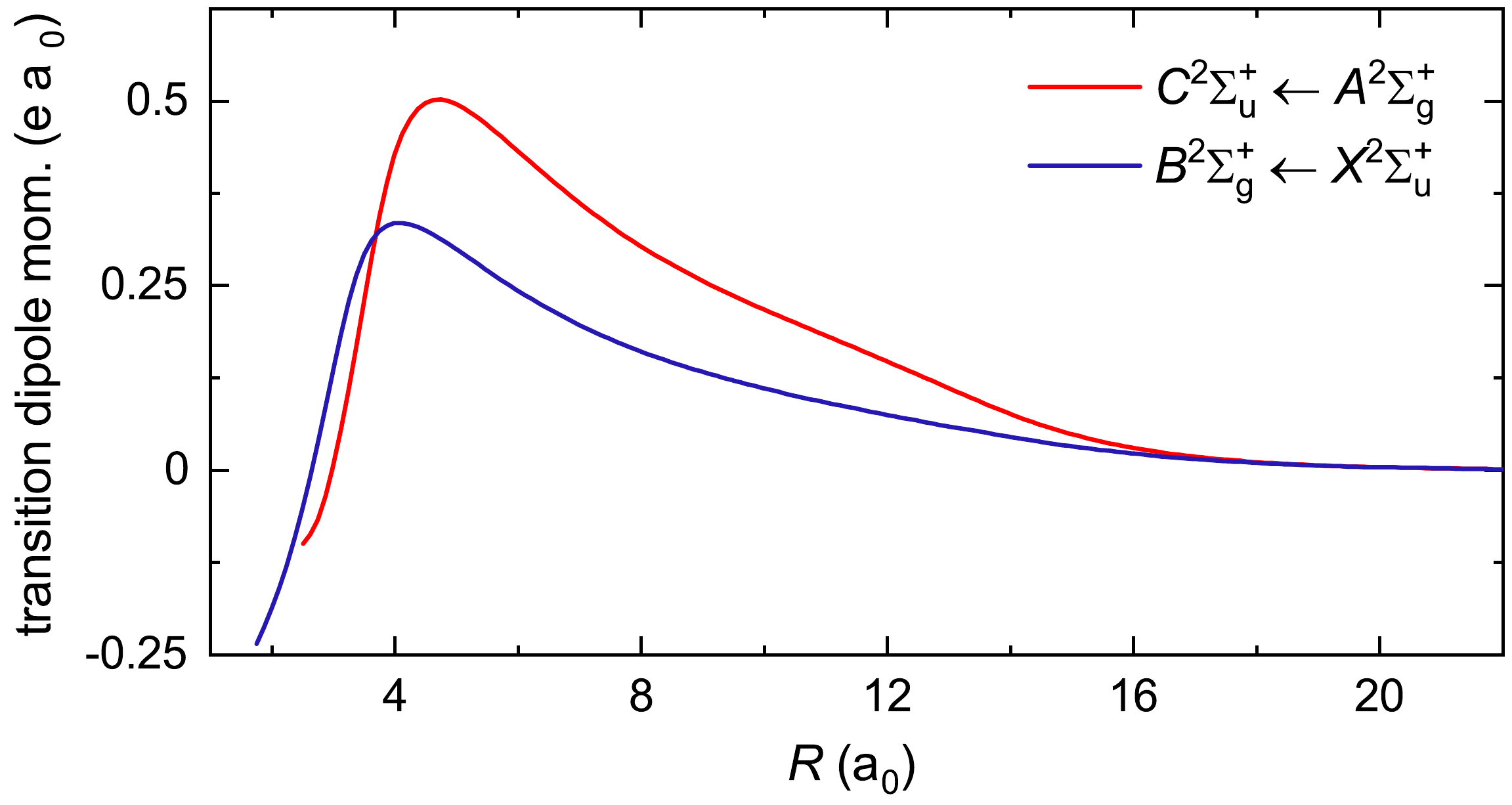}
\end{center}
\caption{The transition electric dipole moments between doublet electronic states of the He$_2^+$ molecular ion.}
\label{fig:DipMom}
\end{figure}

\begin{table*}[tb!]
\caption{The selected transition energies $\hat{\nu}_{\nu J\rightarrow \nu' J'}$ (in cm$^{-1}$) between the rovibrational levels of the He$_2^+$ molecular ion in the ground and excited electronic states. \label{tab:spec3}} 
\begin{ruledtabular}
\begin{tabular}{lcd{7}d{6}d{8}d{7}d{7}d{8}}
State & Isotopologue & \multicolumn{1}{c}{  $\hat{\nu}_{00 \rightarrow 10}\,$} &\multicolumn{1}{c}{   $\hat{\nu}_{00 \rightarrow 20}\,$ } & \multicolumn{1}{c}{  $\hat{\nu}_{00 \rightarrow 02}\,$} &  \multicolumn{1}{c}{ $\hat{\nu}_{00 \rightarrow 04}\,$}  &  \multicolumn{1}{c}{ $\hat{\nu}_{01 \rightarrow 03}\,$} &  \multicolumn{1}{c}{ $\hat{\nu}_{01 \rightarrow 05}\,$} \\
\hline
\multicolumn{8}{c}{He$^+$(${}^2\!S$)+He(${}^1\!S$)} \\
$X^2\Sigma^+_u$ & $^4$He$^4$He$^+$ &1628.4(2) & 3186.5(4) & 42.587(5) & 141.81(2) & 70.937(8) & 198.36(2) \\
			& \multicolumn{1}{r}{ }& &  &  & &  70.93759(6)\footnote{Experimental value from Ref.~\cite{SemeriaPRL20}.} & 198.3647(8)\footnote{\label{note:Semeria}Experimental value from Ref.~\cite{doi:10.1063/1.4967256}.} \\
			& \multicolumn{1}{r}{ }  & 1628.381(3)\footnote{\label{note:FerencPRL20}Theoretical value from Ref.~\cite{FerencPRL20}.}& &  &  & 70.9377(1)\textsuperscript{\ref{note:FerencPRL20}} &  198.364(1)\textsuperscript{\ref{note:FerencPRL20}}  \\
			& $^3$He$^4$He$^+$ &1750.6(2) & 3419.4(5) & 49.486(5) & 164.76(2) & 82.421(9) & 230.42(2) \\
			&  \multicolumn{1}{r}{ } &1750.5554(9)\footnote{Theoretical value from Ref.~\cite{dodatkowyadamowicz}.} & &&&&\\
			& $^3$He$^3$He$^+$ &1863.6(2) & 3634.0(5) & 56.371(6) & 187.65(2) & 93.879(10) & 262.40(3) \\
$A^2\Sigma^+_g$ & $^4$He$^4$He$^+$ &7.285(2) & 8.077(3) & 1.8069(3) & 5.764(1) & 2.9428(5) &   \\
			& $^3$He$^4$He$^+$ &6.999(2) &  & 2.0423(4) & 6.418(1) & 3.3035(6) & \\
			& $^3$He$^3$He$^+$ &6.690(2) & 7.034(3) & 2.2636(4) & 6.972(1) & 3.6316(7) & 	\\
\hline
\multicolumn{8}{c}{He$^+$(${}^2\!S$)+He(${}^3\!S$)} \\
$a^4\Sigma^+_u$	& $^4$He$^4$He$^+$ &311.1(1) & 617.5(3) & 4.2814(4) & 14.267(2) & 7.1344(8) & 19.969(2) \\
			& $^3$He$^4$He$^+$ &335.1(2) & 664.9(4) & 4.9787(5) & 16.590(2) & 8.2962(9) & 23.219(2) \\
			& $^3$He$^3$He$^+$ &357.5(2) & 708.9(4) & 5.6754(6) & 18.911(2) & 9.457(1) & 26.466(3) \\
$B^2\Sigma^+_g$	& $^4$He$^4$He$^+$ & 290.6(7) & 576.7(1.5) & 3.8339(1) & 12.7761(4) & 6.3888(2) & 17.8825(5) \\
			& $^3$He$^4$He$^+$ &313.1(7) & 620.9(1.7) & 4.4582(2) & 14.8562(7) & 7.4291(3) & 20.7931(10) \\
			& $^3$He$^3$He$^+$ &334.0(8) & 662.0(1.8) & 5.0821(3) & 16.934(1) & 8.4684(5) & 23.701(1) \\
$C^2\Sigma^+_u$	& $^4$He$^4$He$^+$ &255.5(1.5) & 507.6(2.4) & 3.132(2) & 10.437(8) & 5.219(4) & 14.61(1) \\
			& $^3$He$^4$He$^+$ &275.3(1.6) & 546.6(2.4) & 3.642(3) & 12.137(10) & 6.069(5) & 16.99(1) \\
			& $^3$He$^3$He$^+$ &293.8(1.6) & 582.9(2.4) & 4.152(3) & 13.83(1) & 6.918(5) & 19.36(2) \\
$b^4\Sigma^+_g$	& $^4$He$^4$He$^+$ &22.873(2) & 43.163(3) & 0.46433(1) & 1.54687(4) & 0.77363(2) & 2.16452(6) \\
			& $^3$He$^4$He$^+$ &24.454(2) & 45.906(3) & 0.53899(1) & 1.79539(5) & 0.89796(2) & 2.51205(7) \\
			& $^3$He$^3$He$^+$ &25.898(2) & 48.378(3) & 0.61338(2) & 2.04297(6) & 1.02183(3) & 2.85820(8) \\
\end{tabular}
\end{ruledtabular}
\end{table*}

\begin{table*}[tb!]
\caption{The numbers of vibrational levels for $J=0$, $N_{\nu,J=0}$, the numbers of all rovibrational levels, $N_{\nu,J}$, and the rotational numbers, $J_\text{max}$, of the most weakly bound rovibrational level of the He$_2^+$ molecular ion in the ground and excited electronic states.  \label{tab:no_of_states}} 
\begin{ruledtabular}
\begin{tabular}{lrrrrrrrrr}
     &\multicolumn{3}{c}{$^4$He$^4$He$^+$}  & \multicolumn{3}{c}{$^3$He$^4$He$^+$} & \multicolumn{3}{c}{$^3$He$^3$He$^+$} \\
\hline
State & $N_{\nu,J=0}$ & $N_{\nu,J}$ & $J_\text{max}$ & $N_{\nu,J=0}$ & $N_{\nu,J}$ & $J_\text{max}$ & $N_{\nu,J=0}$ & $N_{\nu,J}$ & $J_\text{max}$ \\
\hline
\stX & 25  & 832 &  58 &  23  & 710  & 53  & 22  & 626  &  50      \\
\stA& 3  & 9 &  4 &  2  & 7 & 4 &  2  &  7  &   4             \\
\sta& 74 & 5885(1) & 148 & 69  & 5061(4)  & 137 & 65  & 4433(1)  & 128 \\
\stB& 68  & 4953(4)  & 139&  63  & 4259(4) & 129 & 59   &   3732(4) & 120 \\
\stC& 52 & 2833(4) & 113 & 48   & 2429(3) & 105(1) &  45  &2130(3)  & 98  \\
\stb& 19 & 407   &42 & 18   & 350 & 39 & 16   & 306(1)   &36   \\
\end{tabular}
\end{ruledtabular}
\end{table*}

\begin{turnpage}
\begin{table*}[tb!]
\caption{Spectroscopic characteristics of the He$_2^+$ molecular ion in the ground and excited electronic states found using the fit form Eq.~\eqref{Dunham} (constants marked with the prime symbol, e.g., $\omega'_e$), compared with the harmonic constants $\omega_e\, $ and the rotational constants $B_e\, $, which are obtained using the potential energy curves. \label{tab:spec2}}
\begin{ruledtabular}
\begin{tabular}{lcd{9}d{9}d{9}d{9}d{9}d{9}d{9}d{9}d{9}}
State & Isotopologue &\multicolumn{1}{c}{ $\omega'_{e}\,$(cm$^{-1}$)}  & \multicolumn{1}{c}{ $\omega'_e x_e\,$(cm$^{-1}$) }& \multicolumn{1}{c}{ $B'_e\,$(cm$^{-1}$) }& \multicolumn{1}{c}{$\alpha'_e$ (cm$^{-1}$)} & \multicolumn{1}{c}{$\gamma'_e$ $\times 10^{-4}$ (cm$^{-1}$) }&\multicolumn{1}{c}{$\mathcal{D}_e'$ $\times 10^{-4}$ (cm$^{-1}$)  }&\multicolumn{1}{c}{ $\beta_e'$ $\times 10^{-6}$(cm$^{-1}$) }&\multicolumn{1}{c}{  $\omega_e\,$(cm$^{-1}$) }&\multicolumn{1}{c}{  $B_e\,$(cm$^{-1}$)} \\
\hline
\multicolumn{10}{c}{He$^+$(${}^2\!S$)+He(${}^1\!S$)} \\
$X^2\Sigma^+_u$& $^4$He$^4$He$^+$ &	1698.6(2) & 35.11(3) & 7.2128(8) & 0.22304(1) & -13.63(6) & 5.1971(2) & -1.052(1) &	1696.9(2) & 7.2131(8) \\
	&	$^3$He$^4$He$^+$ & 1832.3(3) & 40.86(3) & 8.3922(9) & 0.279804(3)&-18.75(3)  & 7.0339(1) & -1.652(8)& 1830.4(3) & 8.3926(9) \\
	&	 \multicolumn{1}{r}{ }	& & &8.39446(47)\footnote{\label{note:dodad}Theoretical value from Ref.~\cite{dodatkowyadamowicz}.} & 0.28391(15)\textsuperscript{\ref{note:dodad}} & & 7.033(28)\textsuperscript{\ref{note:dodad}} &&& \\
	&	$^3$He$^3$He$^+$ & 1956.8(3) & 46.60(4) & 9.5715(10) & 0.34066(2) &-24.78(1) & 9.14704(5) & -2.45(2) & 1954.8(3) & 9.572(1) \\
\hline
\multicolumn{10}{c}{He$^+$(${}^2\!S$)+He$^*$($^3\!S$)} \\
$a^4\Sigma^+_u$& $^4$He$^4$He$^+$ & 315.79(9) & 2.38(3) & 0.71847(5) & 0.00969(6) & 0.62(7) & 0.1488(1) & 0.157(3) & 315.7(1) & 0.71850(6) \\
	&  $^3$He$^4$He$^+$ & 340.63(9) & 2.77(4) & 0.83596(5) & 0.01215(7)& 0.83(10) & 0.2014(2) &0.228(5)& 340.5(1) & 0.83598(6) \\
	& $^3$He$^3$He$^+$ & 363.8(1) & 3.16(5) & 0.95343(6) & 0.01480(9) & 1.1(1) & 0.2619(3) & 0.316(7) &  363.7(1) & 0.95346(7) \\
$B^2\Sigma^+_g$&   $^4$He$^4$He$^+$ & 295.4(3) & 2.4(2) & 0.6435(2) & 0.0089(4) &  0.9(6) & 0.1224(7)& 0.13(2) & 295.5(3) & 0.6434(1) \\
	&	$^3$He$^4$He$^+$ & 318.6(3) & 2.8(3) & 0.7487(2) & 0.0112(5) & 1.2(8) & 0.166(1)& 0.18(3) & 318.7(4) & 0.7487(2) \\
	&	 $^3$He$^3$He$^+$ & 340.3(3) & 3.2(3) & 0.8539(2) & 0.0136(6) &1.5(10) & 0.216(1)& 0.26(5) & 340.4(4) & 0.8539(2) \\
$C^2\Sigma^+_u$& $^4$He$^4$He$^+$ & 258.9(23) & 1.7(4) & 0.5254(5) & 0.0067(2) & 0.04(9) & 0.0859(7) & 0.07(3) &259.1(22) & 0.5254(4) \\
	& $^3$He$^4$He$^+$ & 279.2(25) & 1.9(5) & 0.6112(6) & 0.0084(2) &-0.02(5) & 0.1161(8) & 0.09(4) & 279.4(23) & 0.6114(5) \\
	& $^3$He$^3$He$^+$ & 298.2(27) & 2.2(6) & 0.6971(7) & 0.0101(3) & -0.11(2) & 0.151(1) & 0.11(4) & 298.4(25) & 0.6973(5) \\
$b^4\Sigma^+_g$& $^4$He$^4$He$^+$ & 25.471(2) & 1.29805(3) & 0.079672(2) & 0.0044737(2) &-1.0334(2) & 0.029673(1) & -0.43355(1) & 25.588(2) & 0.079804(2) \\
	& $^3$He$^4$He$^+$ & 27.474(2) & 1.50853(3) & 0.092704(2) & 0.0056204(2) &-1.3826(2) & 0.039784(2) & -0.65617(2) & 27.601(2) & 0.092855(2) \\
	& $^3$He$^3$He$^+$ & 29.340(2) & 1.71841(3) & 0.105738(3) & 0.0068549(2) &-1.7735(3) & 0.051238(2) & -0.94194(4) & 29.478(2) & 0.105906(3) \\
\end{tabular}
\end{ruledtabular}	
\end{table*}
\end{turnpage}

\begin{table*}[tb!]
\caption{Scattering lengths (in $\text{a}_0$) for collisions of a He$^+$ ion and a He atom in the ground and excited molecular electronic states. \label{tab:spec}} 
\begin{ruledtabular}
\begin{tabular}{lld{6}d{6}d{6}}
State & Potential & \multicolumn{1}{c}{   $^4$He$^4$He$^+$ }& \multicolumn{1}{c}{  $^3$He$^{4}$He$^+$  }& \multicolumn{1}{c}{  $^3$He$^3$He$^+$}  \\
\hline
\multicolumn{5}{c}{He$^+$(${}^2\!S$)+He(${}^1\!S$)} \\
$X^2\Sigma^+_u$	&	$V_\text{BO}$	&	-8.1(7)	&  -45.1(8)	&  85.1(1.4)	\\
			&   $V_\text{BO}$+$V_\text{rel}$	&	-7.9(7)	&  -44.9(8)	& 	85.4(1.4)	\\
			&	$V_\text{BO}$+$V_\text{rel}$+$V_\text{ad}$	&	-8.6(7)	&  -45.8(9)	& 	83.9(1.7)	\\
			&  \ \ theor.\footnote{Scattering lengths calculated using a PEC from Ref.~\cite{TungJCP12}.} & -8.7 & -46.1 &  83.3 \\
$A^2\Sigma^+_g$	&	$V_\text{BO}$	&	328.4(1.1)	&		-142.9(3)& -43.76(6)\\
			&	$V_\text{BO}$+$V_\text{rel}$	&	329.3(1.2)	& 	-142.6(3)	&
			-43.70(4)	\\
			&	$V_\text{BO}$+$V_\text{rel}$+$V_\text{ad}$	&	327.8(1.1)	& 	-142.9(3)	& -43.79(1)	\\
\hline
\multicolumn{5}{c}{He$^+$(${}^2\!S$)+He$^*$(${}^3\!S$)} \\
$a^4\Sigma^+_u$	&	$V_\text{BO}$	& \multicolumn{1}{D{.}{}{6}}{ 	-5500(.800)}& \multicolumn{1}{D{.}{}{6}}{   30(.30)	}& \multicolumn{1}{D{.}{}{6}}{  	3500(.300)}	\\
			&	$V_\text{BO}$+$V_\text{rel}$	& \multicolumn{1}{D{.}{}{6}}{ 	-5000(.700)	}& \multicolumn{1}{D{.}{}{6}}{  	50(.30)	}& \multicolumn{1}{D{.}{}{6}}{ 
			3700(.400)}	\\
			&	$V_\text{BO}$+$V_\text{rel}$+$V_\text{ad}$	& \multicolumn{1}{D{.}{}{6}}{ 	-5900(.900)	}& \multicolumn{1}{D{.}{}{6}}{  20(.30)	}& \multicolumn{1}{D{.}{}{6}}{	3300(.400)	}\\
			
$B^2\Sigma^+_g$	&	$V_\text{BO}$                               & \multicolumn{1}{D{.}{}{6}}{	360(.90)	}& \multicolumn{1}{D{.}{}{6}}{  	230(.70)	}& \multicolumn{1}{D{.}{}{6}}{ 	290(.70)	}\\
			&	$V_\text{BO}$+$V_\text{rel}$               	& \multicolumn{1}{D{.}{}{6}}{	370(.90)	}& \multicolumn{1}{D{.}{}{6}}{ 	250(.80)	}& \multicolumn{1}{D{.}{}{6}}{ 	300(.70)	}\\
			&	$V_\text{BO}$+$V_\text{rel}$+$V_\text{ad}$	& \multicolumn{1}{D{.}{}{6}}{	350(.90)	}& \multicolumn{1}{D{.}{}{6}}{ 	230(.80)	}& \multicolumn{1}{D{.}{}{6}}{ 	280(.70)	}\\
$C^2\Sigma^+_u$	&	$V_\text{BO}$	& \multicolumn{1}{D{.}{}{6}}{	1900(.600)	}& \multicolumn{1}{D{.}{}{6}}{	400(.100)	}& \multicolumn{1}{D{.}{}{6}}{ 		600(.200)	}\\
			&	$V_\text{BO}$+$V_\text{rel}$	& \multicolumn{1}{D{.}{}{6}}{	2000(.600)	}& \multicolumn{1}{D{.}{}{6}}{ 	400(.100)	}& \multicolumn{1}{D{.}{}{6}}{ 	600(.200)	}\\
			&	$V_\text{BO}$+$V_\text{rel}$+$V_\text{ad}$ 	& \multicolumn{1}{D{.}{}{6}}{1900(.600)}& \multicolumn{1}{D{.}{}{6}}{ 		400(.100)	}& \multicolumn{1}{D{.}{}{6}}{ 		600(.200)}\\			
$b^4\Sigma^+_g$	&	$V_\text{BO}$	&87.5(1.6)& \multicolumn{1}{D{.}{}{6}}{ 4290(.30)}& 		\multicolumn{1}{D{.}{}{6}}{ -4740(.30)}\\
			&$V_\text{BO}$+$V_\text{rel}$&93.2(1.6)& \multicolumn{1}{D{.}{}{6}}{  4390(.30)}& 	\multicolumn{1}{D{.}{}{6}}{ -4630(.30)}\\
			&$V_\text{BO}$+$V_\text{rel}$+$V_\text{ad}$&84.9(1.7)& 		\multicolumn{1}{D{.}{}{6}}{ 4250(.40)}& 		\multicolumn{1}{D{.}{}{6}}{ -4800(.90)}\\
\end{tabular}
\end{ruledtabular}
\end{table*}

\subsection{Electric dipole transition moments}

Figure~\ref{fig:DipMom} presents the electric dipole transition moments for the spin-conserving transitions between the doublet electronic states of the He$_2^+$ molecular ion. These transition moments asymptotically decrease to zero in the non-relativistic picture, but their short-range variation may drive the spontaneous interaction-induced deexcitation of a metastable He$^*$($^3\!S$) atom colliding with a He$^+(^2\!S)$ ion, being a limiting factor for the experimental observation and application of such mixtures in the ultracold regime. The \stC~$\leftarrow$~\stA{} transition moment is larger than the \stB~$\leftarrow$~\stX{} one, which means that deexcitation from the \stC{} channel may be faster. The transitions occur at short distances, where the amplitude of the scattering wavefunction is small at ultralow temperatures, which may restrict the rate of this process. Detailed scattering calculations are, however, out of the scope of this work. 

The lowest \sta{} quartet state is stable in the non-relativistic picture, and the decay of the quartet states into the lower doublet molecular states is, in principle, forbidden. The related $^3\!S\leftarrow{}^1\!S$ transition for neutral helium is of the $M$1 type. It is only allowed at $\alpha^3$ order of perturbation theory~\cite{PhysRevA.64.042510} (leading quantum electrodynamics effects), and it is forbidden both at the non-relativistic and leading relativistic levels. The relativistic effects usually are more pronounced at shorter distances and may mix the doublet and quartet electronic states. However, we numerically check that the electric dipole transition moments between the doublet and quartet states are negligible in relativistic calculations with the Dirac-Coulomb Hamiltonian. Thus, the studied quartet states are metastable.

\subsection{Rovibrational structure and spectroscopic constants}

The complete lists of all calculated rovibrational energies for the three stable isotopologues of the He$_2^+$ molecular ion in the ground and excited electronic states are collected in Supplemental Material~\cite{supplemental}. Table~\ref{tab:spec3} collects the selected rotational and vibrational transition energies obtained as a difference between calculated rovibrational energies. We neglect restrictions and selection rules imposed on possible transitions by the bosonic and fermionic symmetries. Our results for the ground electronic state show a good agreement with the previous   experimental~\cite{doi:10.1063/1.4967256,SemeriaPRL20} and theoretical~\cite{FerencPRL20, dodatkowyadamowicz} values within the numerical uncertainties, which indicates that our rovibrational and transition energies should also have similar accuracy for the excited electronic states. The present results agree well with older experimental and theoretical results~\cite{PhysRevLett.115.133202, matyus2018}, too. 

Table~\ref{tab:no_of_states} presents the number of vibrational bound states (rovibrational levels for $J=0$), the total number of rovibrational levels, and the rotational quantum number of the most weakly bound rovibrational level calculated for all stable isotopologues of the He$_2^+$ molecular ion in the ground and excited electronic states. The number of bound states correlates with the well depth and volume of the PEC, where the first-excited weakly-bound electronic state supports just a few levels and higher-excited deeply-bound electronic states support a much larger number of rovibrational levels than the ground state because of their larger volumes and despite their smaller well depths. We find that the number of vibrational states for $J=0$ is well converged for all isotopologues of He$_2^+$, which is necessary to calculate scattering lengths with controlled uncertainties in the next subsection.

Table~\ref{tab:spec2} collects the spectroscopic constants of the He$_2^+$ molecular ion in the ground and excited electronic states calculated by fitting the expression in Eq.~\eqref{Dunham} to the energies of rovibrational levels. Considering a large number of calculated bound states, we use the leave-one-out cross-validation (LOOCV) method to determine the number of rovibrational eigenvalues we should use to fit the expression in Eq.~\eqref{Dunham}. We perform fits using a set of rovibrational energies $E_{\nu,J}$, such that $\nu < \tilde{\nu}$ and $J < \tilde{J}$ for some constraining values $\tilde{\nu}$ and $\tilde{J}$. The LOOCV method allows us to find the optimal values $\tilde{\nu}$ and $\tilde{J}$. We successively remove one value from this set, fit, and then calculate the LOOCV parameter, i.e., the accumulated error between the energies from the fit and the energies from DVR, as defined by the LOOCV method~\cite{MARCHETTI2021107262, CELISSE20082350}. We minimize the value of the LOOCV parameter across many sets of eigenvalues. We determine that the lowest value of the LOOCV parameter is for $\tilde{J} = 3$ and $\tilde{\nu} = 4$ for all molecular electronic states. We omit the spectroscopic coefficients for the \stA{} state, because its shallow PEC supports only 3 vibrational levels for $^4$He$_2^+$. Our fit of the expression in Eq.~\eqref{Dunham} shows the best agreement with 'Fit 2' from Ref.~\cite{dodatkowyadamowicz}. 'Fit 2' assumes the precise value of $\omega_e x_e = 41.1$ taken from Ref.~\cite{wartosc_wexe} and neglects the parameter $\beta_e$. Values from 'Fit 2' are presented for comparison in Tables~\ref{tab:spec3} and~\ref{tab:spec2}. The slight differences between our results and the previous values in Table~\ref{tab:spec2} likely stem from the different number of spectroscopic constants used in the fitted expansion, where the authors of Ref.~\cite{dodatkowyadamowicz} neglected higher vibrational terms. Additionally, Ref.~\cite{dodatkowyadamowicz} includes the non-adiabatic effects, which are missing in our computations. We estimate that the non-adiabatic effects given by Eq.~\eqref{eq:NAE} increase the rotational constant by $7.4\times 10^{-3}\,$\% and $9.5\times 10^{-3}\,$\% for \stX{} and \sta, respectively. 

\subsection{Scattering lengths}
\label{subsec:scattering_lengths}

Accurate PECs are necessary for predicting collisional properties at ultralow temperatures. The most important parameter for ultracold physics experiments is the $s$-wave scattering length, which almost fully characterizes scattering in the ultracold regime but is highly sensitive to the accuracy of the PEC volume and especially to the position of the last weakly bound state~\cite{RevModPhys.82.1225}. Table~\ref{tab:spec} collects the scattering lengths calculated for He$^+$+He collisions in the ground and excited molecular electronic states for three isotopic combinations (neglecting nuclear spins and hyperfine couplings) using our PECs with and without relativistic and adiabatic corrections. The relativistic and adiabatic corrections have little, but not negligible, effect on the scattering lengths. The accuracy of our PECs is high enough to provide the scattering lengths with reasonable uncertainties. Our scattering length for the \stX{} state agrees well with the value derived from the previous accurate PEC~\cite{TungJCP12}. 

\begin{figure}[tb!]
\begin{center}
\includegraphics[width=\columnwidth]{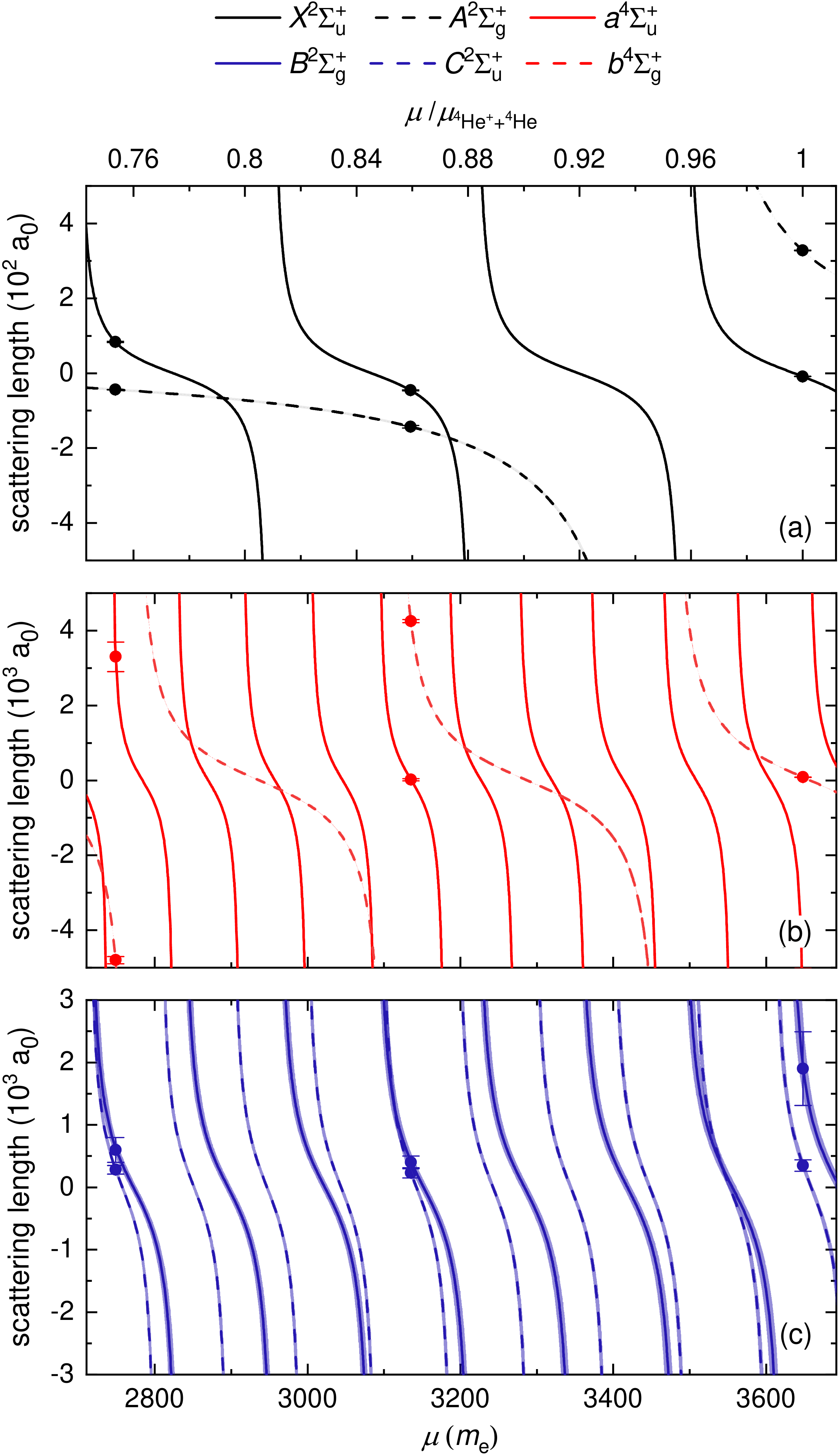}
\end{center}
\caption{The scattering lengths as a function of the reduced mass for the He$^+$+He collisions in the (a)~\stX~and~\stA, (b)~\sta~and~\stb, and (c)~\stB~and~\stC{} molecular electronic states. The points indicate the values for the $^3$He$^3$He$^+$, $^3$He$^4$He$^+$, and $^4$He$^4$He$^+$ isotopologues. The shaded areas (in some cases smaller than the line width) represent the uncertainties of the scattering lengths resulting from the uncertainties of the used PECs.}
\label{fig:as_of_mu}
\end{figure}

The order of magnitude of the scattering lengths can be estimated by the characteristic lengths of interaction, which are $R_4 = 71.0$~a$_0 $ for $^4$He$^+$($^2\!S$)+$^4$He($^1\!S$) and $R^*_4 =  1073.1$~a$_0$ for $^4$He$^+$($^2\!S$)+$^4$He$^*$($^3\!S$). Consequently, several values of the scattering lengths for the higher-excited molecular states are larger than the ones for the ground and first-excited state. In quantum few- and many-body physics, positive and negative signs of the scattering length (both possible for He$^+$+He$^*$ collisions) can be related to the repulsive and attractive effective interactions.  The largest scattering length is for the $^4$He$^+$($^2\!S$)+$^4$He$^*$($^3\!S$) collisions in the \sta{} electronic state, where it is $5.5$ times larger than the characteristic length of interaction. 

The relationship between scattering length, reduced mass, number of bound states, and PEC accuracy can be seen in Fig.~\ref{fig:as_of_mu}, where the scattering lengths for the studied electronic states are plotted as a function of the reduced mass. Faster variation with the reduced mass is visible for deeper PECs supporting larger number of bound states. Large values of some scattering lengths (and their uncertainties) can be explained by the proximity of scattering resonances, which appear when a new vibrational bound state emerges. Large positive scattering lengths are related to the existence of a very weakly-bound vibrational level just below the dissociation limit.

\section{Summary and conclusions}
\label{sec:summary}

We have used state-of-the-art \textit{ab initio} electronic structure methods to calculate the potential energy curves, including the adiabatic and relativistic corrections, for the He$_2^+$ molecular ion in the ground and excited electronic states. We have focused primarily on the higher-excited molecular states resulting from the interaction between a ground-state He$^+$($^2\!S$) ion and a lowest-metastable-state He$^*$($^3\!S$) atom, relevant for proposed cold hybrid ion-atom experiments. The uncertainties of our numerical results have been carefully estimated. We have also reported the transition electric dipole moments. The PECs have been used to calculate and analyze the complete rovibrational structures and spectroscopic constants of three stable isotopologues of the He$_2^+$ molecular ion. The accuracy of our PECs has been high enough to predict and analyze the scattering lengths with controlled and reasonably small uncertainties. Our results for the ground electronic state agree well with previous accurate theoretical and experimental data. The presented numerical approach and analysis can be applied to other three-electron molecular systems such as LiH$^+$, He$^+$+H$_2$, or H$_3$. The calculated potential energy curves, transition dipole moments, and rovibrational energies in a numerical form are collected in the Supplemental Material~\cite{supplemental}.

The reported electronic structure data will be employed in multichannel quantum scattering calculations, including the hyperfine structure to describe ultracold He$^+$($^2\!S$)+He$^*$($^3\!S$) collisions. On the one hand, the rate consonants for interaction- and collision-induced spontaneous deexcitation of metastable atoms and formation of ground-state He$_2^+$ molecular ions can be obtained with the calculated potential energy curves and transition electric dipole moments. On the other hand, the positions and widths of magnetically tunable Feshbach resonances in ultracold He$^+$($^2\!S$)+He$^*$($^3\!S$) mixtures can be predicted to evaluate prospects for their application in quantum physics and chemistry experiments. The presented data can also be used to calculate the radiative lifetimes and absorption spectra of the He$_2^+$ molecular ions.

\

\begin{acknowledgments}
We would like to thank Florian Meinert for inspiring discussions. Financial support from the National Science Centre Poland (Grant No.~2016/23/B/ST4/03231) is gratefully acknowledged. The computational part of this research has been partially supported by the PL-Grid Infrastructure.
\end{acknowledgments}

\appendix*

\section{Adiabatic correction for diatomic systems}

Let us consider a diatomic system AB comprising two monomers (atoms or ions) with nuclei with mass $m_I$ located at positions specified by vectors $\textbf{r}_I$, $I=\text{A},\text{B}$. The adiabatic correction for the molecule, defined as the expectation value of the nuclear kinetic energy operator calculated with the Born-Oppenheimer electronic wavefunction of the system $\Psi$, may be rewritten in the nuclear-center-of-mass coordinate frame as (in atomic units)
\begin{equation}\label{adiab:diatom}
E_\text{ad}(R)=
\frac1{2\mu_n}\left\langle\nabla_\textbf{R}\Psi|\nabla_\textbf{R}\Psi\right\rangle
+\frac1{2M_n}\left\langle\Psi|\textbf{P}^2|\Psi\right\rangle \ ,
\end{equation}
where $\mu_n$ is the reduced mass of the nuclei, $M_n$ is the total mass of the nuclei, $\textbf{R}=\textbf{r}_\text{B}-\textbf{r}_\text{A}$ is the vector joining the nuclei, and $\textbf{P}$ is the total electronic momentum operator, $\textbf{P}=\sum_i\textbf{p}_i$ with $\textbf{p}_i$ being the momentum operator of electron $i$.

To avoid explicit differentiation of the molecular wavefunction $\Psi$ with respect to $\textbf{R}$, we obtain $\nabla_{\textbf{R}}\Psi$ by solving the equation~\cite{NAPT08}
\begin{equation}\label{adiab:diatom:eq}
\Big(H_\text{el}(R)-E_\text{BO}(R)\Big)\nabla_{\textbf{R}}\Psi 
=-\left[\nabla_\textbf{R},H_\text{el}(R)\right]\Psi \ ,
\end{equation}
where $H_\text{el}(R)$ is the clamped-nuclei electronic hamiltonian of the molecule with nuclei separated by the distance $R$, and $E_\text{BO}(R)$ is the electronic BO energy for the molecular state. We construct the right hand side of Eq.~(\ref{adiab:diatom:eq}) with the molecular wavefunction $\Psi$ from the FCI calculations. We then obtain the solution by representing $\nabla_{\textbf{R}}\Psi$ as a separate FCI expansion and solving the set of linear equations for the unknown CI coefficients. Calculation of the second term in Eq.~(\ref{adiab:diatom}) is straightforward, as it is the expectation value of an operator that depends only on electronic coordinates.

The method presented above is exact only if the set of one-electron basis functions used to construct Slater determinants is complete. If the one-electron basis set is incomplete, the effect of neglecting derivatives of basis functions with respect to coordinates of the nucleus they are centered on may be significant. This problem can be alleviated by extending the basis set by adding functions that are derivatives of the functions already present in this set. For example, in the present calculations for the He$_2^+$ system, we found that it is sufficient to augment the orbital basis sets by $p$ functions obtained by taking the nuclear gradient of the contracted $s$ functions representing occupied Hartree-Fock orbitals of helium.

Calculations of the adiabatic correction to the interaction energy, $V_\text{ad}(R)$, performed using incomplete basis sets require also a specific definition of the adiabatic corrections for the monomers. To assure that $V_\text{ad}(R)$ vanishes to zero when $R\to\infty$, the correction for monomer $I=\text{A},\text{B}$ has to be calculated from the formula
\begin{equation}
E_\text{ad}^I=
 \frac{t}{2m_I}\left\langle\nabla_{\textbf{r}_I}\Psi^I|\nabla_{\textbf{r}_I}\Psi^I\right\rangle
+\frac{1-t}{2m_I}\left\langle \Psi^I|\left(\textbf{P}^I\right)^2|\Psi^I\right\rangle,
\end{equation}
where $t=\mu_n/m_I$, and $\Psi^I$ is the wavefunction and $\textbf{P}^I$ is the total electronic momentum operator for the monomer $I$. Explicit differentiation of $\Psi^I$ with respect to the position of the nucleus, $\textbf{r}_I$, is avoided by solving the equation similar to Eq.~\eqref{adiab:diatom:eq}
\begin{equation}
\Big(H_\text{el}^I-E_\text{BO}^I\Big)\nabla_{\textbf{r}_I}\Psi^I
=-\left[\nabla_{\textbf{r}_I},H_\text{el}^I\right]\Psi^I \ ,
\end{equation}
where $H_\text{el}^I$ is the electronic hamiltonian of the monomer, and $E_\text{BO}^I$ is the energy of the state $\Psi^I$. The solution for $\nabla_{\textbf{r}_I}\Psi^I$ is obtained analogously as in the molecular case.

\bibliography{He2+}

\end{document}